\documentclass[preprint]{lmcs}
\pdfoutput=1

\usepackage{lastpage}
\lmcsdoi{20}{3}{18}
\lmcsheading{}{\pageref{LastPage}}{}{}%
{Nov.~15,~2022}{Aug.~29,~2024}{}

\usepackage[utf8]{inputenc}

\usepackage{bussproofs}
\usepackage{fullmacroslmcs}

\theoremstyle{remark}
\newtheorem{notations}[thm]{Notations}
\newtheorem{example}[thm]{Example}

\newcommand{\Fix}[1]{\mathrm{Fix}(#1)}
\newcommand{\measured}[1]{\mathbf{#1}}

\newcommand{\altpath}[1]{\mathrm{AltCycle}(#1)}

\newcommand{\secLL}{\ensuremath{\textsc{ll}^2}\xspace}

\newcommand{\zetameas}[2]{\zeta_{\de{#1},\de{#2}}}
\renewcommand{\cond}[1]{\mathbb{#1}}

\newcommand{\PrimeCPath}[1]{\mathcal{P}[#1]}

\newcommand{\Interpret}[1]{\mathopen{|\![}#1\mathclose{]\!|}}

\newcommand{\amc}{{\sc amc}\xspace}

\newcommand{\Space}[1]{\mathbf{#1}}

\newcommand{\malltwo}{{\sc mall}$^2$\xspace}

\newcommand{\IG}{{\sc ig}\xspace}

\newcommand{\acton}{\curvearrowright}
\newcommand{\distribs}[1]{{\mathbb P}(\measured{#1})}
\newcommand{\nsmp}{{\scshape nsmp}\xspace}

\renewcommand{\identity}[1][]{\mathbf{1}_{#1}}

\renewcommand{\lg}[1]{\mathopen{\vert}#1\mathclose{\rvert}}

\newcommand{\bigre}{{\sc BiGRE}\xspace}
\newcommand{\lobe}{{\sc LoBE}\xspace}
\newcommand{\cohop}{{\sc CoHOp}\xspace}
\newcommand{\dysco}{{\sc DySCo}\xspace}

\begin{document}

\title{Zeta Functions and the (Linear) Logic of Markov Processes}

\author[T. Seiller]{Thomas Seiller\lmcsorcid{0000-0001-6313-0898}}
\address{CNRS, LIPN -- UMR 7030 Université Sorbonne Paris Nord}
\curraddr{99 Avenue Jean-Baptiste Clément, 93430 Villetaneuse, FRANCE}
\email{seiller@lipn.fr}
\thanks{T. Seiller was partially supported by the European Commission Marie Sk\l{}odowska-Curie Individual Fellowship
(H2020-MSCA-IF-2014) project 659920 - ReACT, the INS2I grants \bigre and \lobe, the Ile-de-France DIM RFSI Exploratory project \cohop, and the ANR-22-CE48-0003-01 project \dysco.}

\begin{abstract}
The author introduced models of linear logic known as \enquote{Interaction Graphs} which generalise Girard's various Geometry of Interaction constructions. In this work, we establish how these models essentially rely on a deep connection between zeta functions and the execution of programs, expressed as a cocycle. This is first shown in the simple case of graphs, before being lifted to dynamical systems. Focussing on probabilistic models, we then explain how the notion of graphings used in Interaction Graphs captures a natural class of sub-Markov processes. We then extend the realisability constructions and the notion of zeta functions to provide a realisability model of second-order linear logic over the set of all (discrete-time) sub-Markov processes. 
\end{abstract}

\maketitle
\section*{Introduction}

We construct a mathematical model (semantics) of second-order linear logic using realisability techniques. The standard approach to semantics consists in starting from a logical system to produce a model, which then turns out to capture (well-behaved) programs in a given programming language through the proofs-as-programs correspondence \cite{AJM,hylandong,ehrhard18}. In realisability, one starts from a model of computation and exhibits a logical system arising from it: a type system naturally describing the behaviour of the underlying set of programs. Realisability can then be used to study the relationship between the computational principles used to define programs and the logic of types, such as exhibiting the relationship between bar recursion and the axiom of choice \cite{realizabilitychoice}.

Realisability models are known for intuitionistic and classical logic \cite{realizabilitybook,krivine1,krivine2,miquel}. We are here interested in realisability models for linear logic \cite{seiller-HdR}. While not characterised this way, numerous realisability models were defined in the literature, under different names: ludics \cite{locussolum,ludics,llludintro2,TeruiLudics}, Geometry of Interaction \cite{multiplicatives,towards,goi1,goi2,goi3,feedback,goi5}, Interaction Graphs \cite{seiller-goim,seiller-goiadd,seiller-goig,seiller-goie,seiller-goif}, transcendental syntax \cite{syntran1,syntran2,syntran3,seiller-tilings}. This may be explained by the fact that realisability models for intuitionistic or classical logic are based on the lambda-calculus or variations thereof, while models for linear logic are constructed from varying and less standard models of computation. Recently, the author introduced such models with classes of generalised dynamical systems as the underlying models of computation. This will be formally explained in Sections \ref{section:dynsys} and \ref{section:prob}, where we also show that his construction provides models based on some restricted class of subprobability kernels -- those kernels for which the probability distribution associated to each point is discrete. This raises the question of whether one can extend these ideas to a larger class of kernels, capturing continuous distributions. We will answer positively to this question in section \autoref{section:markovmodel}.

To construct this model, we build on a newfound fundamental property underlying the author's previous constructions. This property is expressed as a \emph{cocycle condition}\footnote{We use this terminology because of the strong resemblance between the condition (Equation \ref{cocycle}) and the notion of $2$-cocycle for a group action.} relating the execution of programs and linear negation. More precisely, we show that the measurement used in Interaction Graphs models to define linear negation corresponds to computing the value at $1$ of Ruelle's zeta function for dynamical systems \cite{Ruellezeta}. This is then used as a guideline: we define a \emph{zeta kernel} for subprobabilistic kernels which is then used to define a realisability model for second-order linear logic.

\paragraph{Contributions and plan of the paper}

The main and more salient contribution of this work is the definition of a model of second-order linear logic (\secLL) from realisability techniques applied to general sub-Markov kernels. This is, to the author's knowledge, the first model of \secLL able to accommodate discrete and continuous probability distributions. This opens the possibility of defining new models of typed lambda-calculus extended with probabilities and specific instructions for sampling discrete and non-discrete distributions. This result is guided by another, more technical, contribution establishing that previous linear realisability models were defined upon a formal geometric connection between program execution and zeta functions (for graphs and dynamical systems). The second major contribution of this work, one which opens many research directions for future work, is the discovery of this fundamental connection between program execution and zeta functions.

The obtention of these results goes through several steps, each of which consists of a separate contribution.
\begin{itemize}
\item In Section \ref{section:graphs}, we recall the author's discrete Interaction Graphs (\IG) models based on graphs and relate them with the work of Ihara on zeta functions of graphs \cite{IharaZeta}. This section is mostly introductory: we explain in the simple setting of graphs how the models of linear logic (in this case only the multiplicative fragment \MLL) are constructed, and in particular how the notion of type is inferred. We however already state the first contribution of this work: showing how these models essentially rely on a cocycle relation involving the notions of execution and Ihara's graph zeta functions.
\item In Sections \ref{section:dynsys} and \ref{section:prob}, we show how this fundamental observation lifts to the more involved models based on \emph{graphings} \cite{seiller-goig,seiller-goif}. We first recall the basic notions and prove that the restriction to \emph{deterministic} graphings boils down to a representation of programs as partial dynamical systems, showing how the Interaction Graphs constructions provide realisability models for linear logic on dynamical systems. We then prove that in this more general case, the models once again rely on a fundamental cocycle involving execution (here related to the iteration of the dynamical system) and Ruelle's zeta function for dynamical systems. Similar results are obtained for the notion of \emph{probabilistic graphings} which we show coincide with a specific class of sub-Markov processes that we describe. 
\item In Sections \ref{section:markovzeta} and \ref{section:markovmodel}, we generalise the realisability constructions from the previous sections to the set of all sub-Markov kernels. We therefore introduce the notions of execution and zeta functions for general sub-Markov kernels and prove they satisfy the essential cocycle relation. These results are then used to define realisability models of second-order linear logic over the set of all sub-Markov kernels.
\end{itemize}

\paragraph{Related work}

The results presented here are strongly related to the Geometry of Interaction program of Girard \cite{multiplicatives,towards,goi1,goi2,goi3,locussolum,feedback,goi5,seiller-goim,seiller-goiadd,seiller-goig,seiller-goie,seiller-goif,syntran1}. As a consequence, it is close in spirit to game semantics approaches to probabilistic programs \cite{danosprobgames,paquetprob1,paquetprob2,paquetprob3,paquetprob4}. 
There are strong connections between Geometry of Interaction semantics and coherence space semantics for linear logic. One might expect some functorial relationship (maybe based upon the antisymmetric tensor algebras -- or \emph{Fock} -- functor) between the work presented here and either probabilistic coherence spaces \cite{qcs,ehrhard11,ehrhard14,ehrhard18,ehrhard19}, or denotational semantics related to stochastic kernels \cite{girardbanach,kerjean18,kerjean19,panangaden1}.

\section{Interaction Graphs: the discrete case \label{section:graphs}}

In this section, we review the models of linear logic introduced by the author under the name \enquote{Interaction Graphs}. While the next section will be devoted to the general case where programs are represented as \emph{graphings}, in this section we restrict to the more simple specific case of programs represented as graphs. 

Graphs provide a minimal but natural mathematical structure to represent programs. Indeed, Turing machines and automata can naturally be abstracted as finite graphs. Obviously, some information is lost by considering discrete graphs: following an edge in a transition graph corresponds to performing some instruction which modifies the state of the machine, something that cannot be accounted for with finite graphs. Restricting to finite structures in some sense limits the approach to models of computation with finite sets of states. To regain expressivity, the author introduced graphings \cite{seiller-goig}, which will be formally defined in Section \ref{sec:graphings}. Graphings are graphs realised on a topological or measured space which represent the space of all possible configurations of the machine. This allows us to interpret edges of the transition graph as specific endomorphisms of this space, recovering the expressivity lost by considering only discrete structures. 

In order to ease the presentation, we restrict the discussion to finite graphs in this section, stressing that all the intuitions built in this easier setting will carry over in the next sections. In this model for which a program is abstracted as a finite graph, computation is represented as the formation of paths in the graph. In the case of Turing machines the graph represents the transition function, and the process of computation corresponds to the iteration of this transition function, i.e. following a path to travel through the graph -- a sequence of instructions. The dynamic process of computation itself is therefore represented as the operation of \emph{execution} $\mathrm{Ex}(G)$ of a graph $G$, which is the set of maximal paths in $G$. This alone describes some kind of abstract, untyped, model of computation, which one can structure by defining types depending on how graphs behave. From the point of view of logic, this operation of execution computes the normal form of a proof, i.e. accounts for the cut-elimination procedure.

Types are then defined as sets of graphs (satisfying some properties), i.e. a type $\cond{A}$ is identified with the set of all programs typable by $\cond{A}$. The notion of \emph{execution}, which abstractly represents the execution of a program given some input, is the key ingredient to the construction of (linear) implication, i.e. arrow types. Indeed, supposing $\cond{A}, \cond{B}$ are defined types, then a graph $G$ will have type $\cond{A\multimap B}$ (the linear implication) if and only if for all graphs $A$ of type $\cond{A}$, the graph $G$ applied to $A$ -- noted $G\bicol A$ -- reduces to a graph $\mathrm{Ex}(G\bicol A)$ of type $\cond{B}$. Let us note that this formalism is extremely expressive. For instance it naturally interprets polymorphism (a graph belongs to many sets of graphs, thus many types), subtyping (the inclusions of sets of graphs), and quantifiers (defined through unions and intersections of sets of graphs). 

One key observation is that a type is not just any set of programs, but one satisfying a closure property. More specifically, a type is a set $\cond{A}$ of graphs such that $\cond{A}=\cond{A}^{\poll{}{}\poll{}{}}$, where $\poll{}{}$ is an \emph{orthogonality relation} accounting for linear negation. Equivalently, a type is a set $\cond{A}$ of graphs such that $\cond{A}=T_\cond{A}^{\poll{}{}}$ for some set $T_\cond{A}$ understood as a set of \emph{tests}. For instance, the natural tests for a graph $F$ of type $\cond{A\multimap B}$ consist of pairs $(A,B')$ where $A$ is an element of $\cond{A}$ given as input and $B'\in\cond{B}^{\poll{}{}}$ is used to test the result of the computation can be given the type $\cond{B}$. From the point of view of logic, this interactive view of the definition of linear negation extends the notion of \emph{correctness criterion} for \MLL proof nets. 

After giving this informal overview, we will formally define the models based on graphs. We first review basic definitions, and then explore the relationship with Ihara's zeta function of a graph, which will be extended to the more general setting of graphings in \autoref{section:dynsys}. 

\subsection{Interaction Graphs: basic notions}

We briefly recall the basics of the Interaction Graphs (\IG) model in the discrete case. We work with weighted directed (multi-)graphs; here we will suppose weights are picked in the field of complex numbers $\complexN$. Graphs are defined as tuples $G=(V^G,E^G,s^G,t^G,\omega^G)$, where $V^G$ and $E^G$ are sets, $s^G$ and $t^G$ are respectively the source and target maps from $E^G$ to $V^G$, and $\omega^G: E^G\rightarrow \complexN$ is a weight map. 

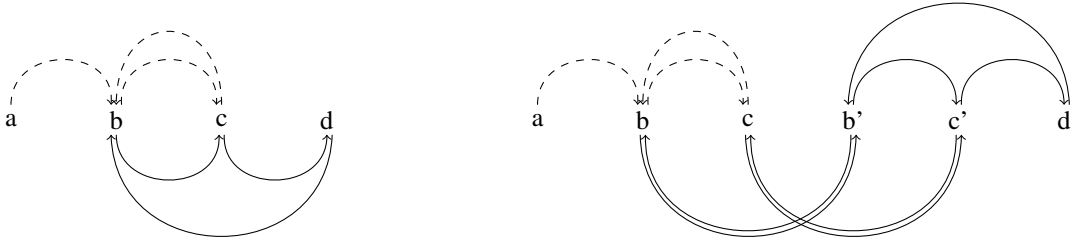
\begin{figure*}
\centering
\begin{tabular}{cc}
\begin{tikzpicture}[x=0.7cm]
\node (a) at (0,0) {a};
\node (b) at (2,0) {b};
\node (c) at (4,0) {c};
\node (d) at (6,0) {d};

\draw[->,dashed] (0,0.2) .. controls (0,1) and (2,1) .. (1.9,0.2) {};
\draw[->,dashed] (2.1,0.2) .. controls (2,1) and (4,1) .. (3.9,0.2) {};
\draw[->,dashed] (4,0.2) .. controls (4,1.5) and (2,1.5) .. (2,0.2) {};

\draw[<-] (6,-0.2) .. controls (6,-1) and (4,-1) .. (4.05,-0.2) {};
\draw[->] (2,-0.2) .. controls (2,-1) and (4,-1) .. (3.95,-0.2) {};
\draw[->] (6.1,-0.2) .. controls (6,-2) and (2,-2) .. (1.9,-0.2) {};

\node (a') at (10,0) {a};
\node (b') at (12,0) {b};
\node (c') at (14,0) {c};
\node (b'') at (16,0) {b};
\node (c'') at (18,0) {c};
\node (d') at (20,0) {d};

\draw[->,dashed] (10,0.2) .. controls (10,1) and (12,1) .. (11.9,0.2) {};
\draw[->,dashed] (12.1,0.2) .. controls (12,1) and (14,1) .. (13.9,0.2) {};
\draw[->,dashed] (14,0.2) .. controls (14,1.5) and (12,1.5) .. (12,0.2) {};

\draw[<-] (20,0.2) .. controls (20,1) and (18,1) .. (18.05,0.2) {};
\draw[->] (16,0.2) .. controls (16,1) and (18,1) .. (17.95,0.2) {};
\draw[->] (20.1,0.2) .. controls (20,2) and (16,2) .. (15.9,0.2) {};

\draw[->] (11.95,-0.2) .. controls (12,-2) and (16,-2) .. (16.05,-0.2) {};
\draw[<-] (12.05,-0.2) .. controls (12,-1.9) and (16,-1.9) .. (15.95,-0.2) {};
\draw[->] (13.95,-0.2) .. controls (14,-2) and (18,-2) .. (18.05,-0.2) {};
\draw[<-] (14.05,-0.2) .. controls (14,-1.9) and (18,-1.9) .. (17.95,-0.2) {};

\end{tikzpicture}
\end{tabular}
\caption{On the left: implicit cut between two graphs (one is plain, the other is dashed). On the right: explicit cut between the same two graphs (the cut is shown below). Both representations lead to the same result, as the cut-elimination is represented by \emph{execution}, an operation defined from alternating paths. It is easily checked that there is a bijective correspondence between alternating paths on the left and alternating paths on the right.}\label{executioncut}
\end{figure*}

The first essential operation is that of \emph{execution} between two graphs $F,G$. This interprets program execution (explaining the naming convention) through cut-elimination. The cut is implicitly represented as the common vertices of the two graphs $F,G$. This eases the expressions, and is equivalent to the more traditional approach where one would consider both $F$ and $G$, together with a graph representing the cut rule (cf. Figure \ref{executioncut}). As execution is defined through alternating paths, the results are equivalent and we urge the reader to use whatever convention she finds more natural.

We start to fix a few notations that will be used in this paper.

\begin{notations}
Given two sets $A,B$, we write $A\backslash B$ for the set $\{ a\in A\mid a\not\in B\}$, and $A\triangle B$ for their symmetric difference $(A\backslash B)\cup (B\backslash A)=(A\cup B)\backslash(A\cap B)$.
\end{notations}

\begin{notations}
Given two graphs $G,H$, we write $G\cup H$ for the graph $(V^G\cup V^H, E^G\disjun E^H, s^G\disjun s^H, t^G\disjun t^H, \omega^G\disjun \omega^H)$. Note the non-disjoint union of sets of vertices, which is essential to consider alternating paths between the two graphs.
\end{notations}

\begin{defi}[Alternating paths]
Let $G$ and $H$ be two graphs. An \emph{alternating path} $\pi$ of length $\lg{\pi}=k$ between $G$ and $H$ is a path $(e_{i})$ in $G\cup H$ which satisfies that for all $i=0,\dots,k-1$, $e_i\in E^G$ if and only if $e_{i+1}\in E^H$. The source and target of the path are respectively defined as $s^{G\cup H}(\pi)=s^{G\cup H}(e_0)$ and $t^{G\cup H}(\pi)=t^{G\cup H}(e_{k-1})$.

The set of alternating paths will be denoted by $\enpaths{G,H}$, while $\enpaths{G,H}_{V}$ will mean the subset of alternating paths between $G$ and $H$ with source and target in $V$.
\end{defi}

\begin{defi}\label{definitionExecution}
Let $G$ and $H$ be two graphs. The \emph{execution} of $G$ and $H$ is the graph $G\plug H$ defined by:
\begin{equation*}
\begin{array}{l}
V^{G\plug H}=V^{G}\triangle V^{H},
\hspace{4em}
\!\! E^{G\plug H}=\enpaths{G,H}_{V^{G}\triangle V^{H}}\\
s^{G\plug H}=\pi\mapsto s^{G\cup H}(\pi),
\hspace{2em}
t^{G\plug H}=\pi\mapsto t^{G\cup H}(\pi)\\
\omega^{G\plug H}= \pi=\{e_{i}\}_{i=0}^{n}\mapsto \prod_{i=0}^{n}\omega^{G\bicol H}(e_{i})
\end{array}
\end{equation*}
\end{defi}

This notion of execution can be related to cut-elimination in proof nets, and it represents the execution of programs through the Curry-Howard correspondence. We will now define the notion of orthogonality which can be related to correctness criteria for proof nets, and is used to define types by means of \emph{testing}. We refer the interested reader to work by Naibo, Petrolo and Seiller \cite{seiller-axioms} for more details and explanations. Defining orthogonality in \IG models is done by quantifying \emph{closed paths} and \emph{prime closed paths}.

\begin{defi}
Given a graph $G$, a closed path $\pi$ (called a \emph{circuit} in earlier work \cite{seiller-goim}) of length $\lg{\pi}=k$ is a path $(e_i)_{i=0}^{k-1}$ such that $s^G(e_0)=t^G(e_{k-1})$ and considered up to cyclic permutations. A \emph{prime closed path} (called a $1$-circuit in \IG) is a closed path which is not a proper power of a smaller closed path. We denote by $\uncircuits{G}$ the set of prime closed paths in $G$.
\end{defi}

\begin{defi}
Given graphs $G,H$, an \emph{alternating closed path} $\pi$ of length $\lg{\pi}=2k$ is a closed path $(e_{i})_{0\leqslant i\leqslant 2k-1}$ in $G\cup H$ such that for all $i\in \mathbf{Z}/2k\mathbf{Z}$, $e_i\in G$ if and only if $e_{i+1}\in H$. The set of prime alternating closed paths between $G$ and $H$ will be denoted $\uncircuits{G,H}$.
\end{defi}

This notion is used in previous Interaction Graphs (\IG) models to define a \emph{measurement} which in turn defines the orthogonality relation. The orthogonality is the essential ingredient to define types using realisability techniques. We only recall the measurement here and refer to the first \IG paper for more details \cite{seiller-goim}. The notion of measurement depends on a map that is used to associate each cycle with a positive real number depending on its weight. We will write $\realposN$ the set of non-negative real numbers in the following.

\begin{defiC}[{\cite[Definition 14]{seiller-goim}}]\label{definitionMeasurement}
Let $m$ be a map $\complexN\rightarrow \realposN$. For any two graphs $G,H$ we define the measurement 
\[ \meas{G,H}=\sum_{\pi\in\uncircuits{G,H}} m(\omega^{G\cup H}(\pi)). \]
\end{defiC}

Based on these two ingredients (execution and measurement), and two essential properties, namely the associativity of execution \cite{seiller-goim} and the \emph{trefoil property} \cite{seiller-goiadd}, one can define a myriad of models of Multiplicative Linear Logic (\MLL) and Multiplicative-Additive Linear Logic (\MALL). As shown by the author, these models capture all the different Geometry of Interaction models introduced by Girard by choosing carefully the map $m$ used to define the measurement \cite{seiller-goiadd}. We will now explain how there is a similarity between the measurement just recalled, and the Bowen-Lanford zeta function of graphs. To formalise the connection, we need to consider zeta functions of weighted graphs, but we will start with a quick overview of the theory of zeta functions of (non-weighted) graphs. This connection will then be used to define \IG models of multiplicative linear logic.

\subsection{Bowen-Lanford Zeta Functions}

We first recall the definition and some properties of the zeta function of a directed graph. We refer to the book of Terras \cite{terras2010zeta} for more details. We will later on continue with zeta functions for weighted directed graphs, and further with zeta functions for dynamical systems. The graph case is important as it provides intuitions about the later generalisations.

In this subsection only, we consider non-weighted directed graphs (i.e. there is no weight map $\omega^G$ or, equivalently, this map is the constant map equal to $1$) and suppose they are \emph{simple}, i.e. that the map $E^G\mapsto V^G\times V^G; e\mapsto (s^G(e),t^G(e))$ is injective. Given such a graph, its \emph{transition matrix} is defined as the $V^G\times V^G$ matrix whose coeficients are defined by $M_G(v,v')=1$ if there is an edge $e\in E^G$ such that $s^G(e)=v$ and $t^G(e)=v'$, and $M_G(v,v')=0$ otherwise.
The following definition provides a clear parallel with the famous Euler zeta function.

\begin{defi}
The Bowen-Lanford zeta function associated with the graph $G$ is defined as:
\begin{equation}
\zeta_{G}(z) = \prod_{\tau\in\uncircuits{G}} (1-z^{\lg{\tau}})^{-1} 
\end{equation}
which converges provided $\abs{z}$ is sufficiently small.
\end{defi}

The two following lemmas are easy to establish (using that $-\log(1-x)=\sum_{k=1}^{\infty}\frac{x^n}{n}$). The first is essential in our work, as it provides an alternative expression of the zeta function that we will be able to generalise in \autoref{section:dynsys} and \autoref{section:prob}. Indeed, while the formal definition above uses the notion of prime closed paths, this one quantifies over all closed paths.

The second lemma is key to the representation of $\zeta_G(z)$ as a rational function. This relates the zeta function with the determinant of the adjacency matrix of $G$. Notice that this relation was obtained by the author in the special case $z=1$ \cite{seiller-goim} and was the initial motivation behind the definition of orthogonality in Interaction Graphs models, since it relates the measurement with the Fuglede-Kadison determinant of operators \cite{FKdet} used in Girard's model \cite{goi5}.

\begin{lem} Let $N(n)$ denote the number of all possible strings $(v_1,\dots, v_n)$ representing a closed path in $G$ of length $n$. Then:
\begin{equation}
\zeta_G(z)=\exp\left(\sum_{n=1}^{\infty} \frac{z^n}{n}N(n)\right).
\end{equation}
\end{lem}

\begin{lem}
Let $G$ be a graph and $M(G)$ its \emph{transition matrix}, then $\tr(M(G)^k) = N(k)$.
\end{lem}

The previous lemmas are standard results from the theory of Zeta functions. Together, they yield the following result.

\begin{prop}\label{logdetgraph}
Let $G$ be a graph and $M(G)$ its \emph{transition matrix}. Then:
\[  \log(\zeta_{G}(z)) = 	-\log(\det(1-z.M(G))), \]
for sufficiently small values of $\abs{z}$.
\end{prop}

\begin{proof}
We compute:
\begin{eqnarray*}
\log(\zeta_{G}(z)) 	&=&	\sum_{n=1}^{\infty} \frac{z^n}{n}N(n)\\
 				&=& 	\sum_{n=1}^{\infty} \frac{z^n}{n}\tr(M(G)^n)\\
				&=&	\sum_{n=1}^{\infty} \frac{\tr((z.M(G))^n)}{n}\\
				&=& 	-\log(\det(1-z.M(G))),
\end{eqnarray*}		
where the last equality is found with a (simple) proof in earlier work \cite[Lemma 61]{seiller-goim}.
\end{proof}

As we will show later on, the zeta function of graphs is strongly related to the orthogonality in \IG models, as the measurement used in these models boils down to computing the value of some graph zeta function at $z=1$. In fact, we will show how to define new models by simply considering the zeta function itself instead of its value at $1$. But for this we need to define the zeta function of weighted graphs.

\subsection{Zeta functions of weighted directed graphs}

Now, we consider weighted directed graphs, i.e. graphs with weights of the edges, and we will restrict to the case of complex numbers as weights. We write $\omega$ the weight function, as well as its extension to paths, using the product, i.e. \[\omega(\pi)=\prod_{e\in\pi} \omega(e).\]

Similarly to the case of unweighted graphs, we define the \emph{transition matrix} of a \emph{simple} weighted graph as the $V^G\times V^G$ matrix with $M_G(v,v')=\omega(e)$ if there exists a (necessarily unique) edge $e\in E^G$ with $\langle s^G(e), t^G(e)\rangle =(v,v')$, and $M_G(v,v')=0$ otherwise. 

For a general (i.e. non-simple) weighted graph $G$, we write $G(v,v')$ for the set $\{ e\in E^G\mid s^G(e)=v, t^G(e)=v'\}$. One can then extend the definition of \emph{transition matrix} by associating to $G$ the $V^G\times V^G$ matrix with $M_G(v,v')=\sum_{e\in G(v,v')} \omega(e)$. Alternatively, this matrix can also be defined as $M_{\hat{G}}$ where $\hat{G}$ is the \emph{simple collapse} of $G$, i.e. the simple graph defined as $\hat{G}=(V^{G},\hat{E}^G,\hat{s}^G,\hat{t}^G,\hat{\omega}^G)$ with:
\begin{itemize}
\item $\hat{E}^G=\{(v,v')\in V^G\times V^G\mid G(v,v')\neq\emptyset\}$,
\item $\hat{s}^G((v,v'))=v$,
\item $\hat{t}^G((v,v'))=v'$,
\item $\hat{\omega}^G((v,v'))=\sum_{e\in G(v,v')} \omega(e)$.
\end{itemize}
We here note that the author proved in earlier work (in the special case of graphs with weights in $[0,1]$) that the measurement defined from the function\footnote{In this paper, we use the lambda notation to define  functions, i.e. $\lambda x. \log(1-x)$ denotes $x\mapsto \log(1-x)$.} $m:= \lambda x. \log(1-x)$ satisfies $\meas{F,G}=\meas{\hat{F},\hat{G}}$ \cite[Proposition 16]{seiller-goim}.

The zeta function of a weighted graph is defined as follows.

\begin{defi}
The zeta function associated with the weighted graph $G$ is defined as:
\[  \zeta_{G}(z) = \prod_{\pi\in\uncircuits{G}} (1-\omega(\pi).z)^{-1} \]
which converges provided $\abs{z}$ is sufficiently small.
\end{defi}

Readers familiar with zeta functions of weighted graphs will notice that we take the product of the weights to define the weight $\omega$ of a path, while standard work on zeta functions for weighted graphs define the weight $\nu$ of a path as a sum. This is formally explained by taking a logarithm, i.e. $\nu=\log\circ\omega$, explaining why we here multiply expressions $1-\omega(\pi)z$ instead of $1-z^{\nu(\pi)}$ in the standard definition.

Adapting the proof of the non-weighted case (Proposition \ref{logdetgraph}), one obtains the following general result, which extends the author's combinatorial interpretation of the determinant $\det(1-M(G))$ \cite[Corollary 61.1]{seiller-goim}.
\begin{prop}
Let $G$ be a directed weighted graph and $M(G)$ its \emph{transition matrix}. Then:
\[  \log(\zeta_{G}(z)) = 	-\log(\det(1-z.M(G))), \]
for sufficiently small values of $\abs{z}$.
\end{prop}

Taking the logarithm we obtain:
$$ \log(\zeta_{G}(z)) = \sum_{\pi\in\uncircuits{G}} -\log(1-\omega(\pi).z)), $$
an expression that appears in the definition of measurement in the previous section.
In the restricted case when weights are taken as elements of $]0,1]$, this can be used to relate the measurement defined in Interaction Graphs for $m:=\lambda x.\log(1-x)$ with the value of the zeta function at $z=1$:
$$ \meas[\lambda x.\log(1-x)]{F,G} = \log(\zeta_{F\bullet G}(1)) $$
where the $\bullet$ operation consists in composing (i.e. taking length-2 paths) the graphs $F\cup\emptyset_{V^{F}\backslash V^{G}}$ and $G\cup\emptyset_{V^{G}\backslash V^{F}}$, with $\emptyset_{V}$ the graph $(V,\emptyset,\emptyset,\emptyset)$.

Orthogonality in \IG models is defined by $F\poll{}{} G$ if and only if $\meas{F,G}\neq 0$ or $\infty$, i.e. $-\log(\zeta_{F\bullet G}(1))\neq 0$ or $\infty$. Through this previous result, this is equivalent to the fact that $\zeta_{F\bullet G}(1)\neq 0,1$. We will now build on this remark to extend the construction of \IG models. This provides a new family of models using zeta functions to define the orthogonality.

\subsection{Zeta, Execution and a Cocycle Property}

As we mentioned earlier, there are two essential properties that ensure that \IG realisability models represent (multiplicative additive) linear logic \cite{seiller-goiadd,seiller-phd}. The first is the associativity of execution
\begin{equation} 
F\plug (G\plug H)=(F\plug G)\plug H. \label{associativity}
\end{equation} 
The second is the so-called \emph{trefoil property} \cite{seiller-goiadd}:
\begin{equation} \meas{F,G\plug H}+\meas{G,H} = \meas{G,H\plug F}+\meas{H,F}. \label{trefoilppty}
\end{equation}

Those properties are satisfied under some mild hypothesis on the graphs (i.e. that $V^F\cap V^G\cap V^H=\emptyset$). Technically speaking, the trefoil property is obtained as a consequence of a geometric identity \cite{seiller-phd,seiller-goiadd} establishing that when $V^{F}\cap V^{G}\cap V^{H}=\emptyset$, there is weight-preserving bijection between the following sets of closed paths:
\begin{equation}
\uncircuits{F,G\plug H}\disjun\uncircuits{G,H}\equiv\uncircuits{G,H\plug F}\disjun\uncircuits{H,F}. \label{geometrictrefoil}
\end{equation}

This geometric identity can be used to rephrase \autoref{trefoilppty} as a special case of a general cocycle condition satisfied by zeta functions. Indeed, using the fact we noticed earlier that $\meas[\lambda x.\log(1-x)]{F,G}=\log(\zeta_{F\bullet G}(1))$, the trefoil property (\autoref{trefoilppty}) is a straightforward consequence of the following theorem (when taking $z=1$).

\begin{thm}\label{cocycle1}
Suppose $V^{F}\cap V^{G}\cap V^{H}=\emptyset$. Then:
\begin{equation}
\zeta_{F\bullet (G\plug H)}(z).\zeta_{G\bullet H}(z)=\zeta_{G\bullet (H\plug F)}(z).\zeta_{H\bullet F}(z). \label{cocycle}
\end{equation}
\end{thm}

\begin{proof}
By definition and the geometric trefoil property:
\begin{align*}
\zeta_{F\bullet(G\plug H)}(z).\zeta_{G\bullet H}(z) 
&= \prod_{\pi\in\uncircuits{F,G\plug H}}(1-\omega(\pi).z)^{-1} \prod_{\pi\in\uncircuits{G, H}}(1-\omega(\pi).z)^{-1}\\
&=  \prod_{\pi\in\uncircuits{F,G\plug H}\disjun\uncircuits{G,H}}(1-\omega(\pi).z)^{-1}\\
&=  \prod_{\pi\in\uncircuits{G,H\plug F}\disjun\uncircuits{H,F}}(1-\omega(\pi).z)^{-1}\\
&=  \prod_{\pi\in\uncircuits{G,H\plug F}}(1-\omega(\pi).z)^{-1}\prod_{\pi\in\uncircuits{H,F}}(1-\omega(\pi).z)^{-1}\\
&=\zeta_{G\bullet (H\plug F)}(z).\zeta_{H\bullet F}(z)
\end{align*}
which is what we wanted to prove.
\end{proof}

We can then define families of models of linear logic extending the Interaction Graphs approach by considering the following constructs. We change the terminology w.r.t. earlier papers to avoid conflicts. We use the term \emph{proof-object} in place of the term \emph{project}, and we call \emph{types} what was called a \emph{conduct}. We also use the term \emph{antipode} for the set of functions defining the orthogonality relation, as the standard term \enquote{pole} might be confused with the notion of pole from complex analysis.

Following the constructions from previous papers, we will interpret proofs by pairs $(g,G)$ of a graph $G$ and a function $g$ (which we may call the \emph{wager} following previous terminology) capturing the extra terms $\zeta_{H\bullet F}(z)$ and $\zeta_{G\bullet H}(z)$ in the above property. This technical trick allows us to derive the adjunction underlying the duality of the multiplicative connectives (i.e. ensuring that $(\cond{A\otimes B^{\pol}})^{\pol} =\cond{A\multimap B}$). For more detailed explanations on the wager, and how to recover the adjunction from the above trefoil property, we refer to previous papers on Interaction Graphs \cite{seiller-goim,seiller-goiadd}.

\begin{defi}
A \emph{proof-object} of support $V$ is a pair $(g,G)$ of a function $g:\complexN\rightarrow\complexN$ and a graph $G$ with $V^G=V$.
\end{defi}

\begin{defi}
Given two proof objects $\de{g}=(g,G)$ and $\de{h}=(h,H)$ we define the \emph{zeta-measurement} as the partial complex function:
$ \zetameas{g}{h}=g\cdot h\cdot \zeta_{G\bullet H}$,
where $\cdot$ denotes the pointwise multiplication of functions.
\end{defi}

\begin{defi}
An antipode $P$ is a family of functions $\complexN\rightarrow\complexN$. Given two proof objects $\de{g}=(g,G)$ and $\de{h}=(h,H)$, they are orthogonal w.r.t. the antipode $P$ -- denoted $\de{g}\poll[P] \de{h}$ -- if and only if $\zetameas{g}{h}\in P$.
Given a set $E$ of proof objects, we define its orthogonal as $E^{\poll[P]}=\{\de{g}\mid \forall \de{e}\in E, \de{e}\poll[P]\de{g}\}$.
\end{defi}

We note that many interesting properties of the graph can be used to define orthogonality in this case. Indeed, a number of properties (e.g. connectedness) and invariants (e.g. Euler characteristic) of a graph can be related to analytic properties of the zeta function of a graph. We also note that previous notions of orthogonality \cite{seiller-goim} can be recovered by considering as antipode the set of functions $f$ such that $f(1)\neq 0,1$.

We now define types and explain how models of \MLL can be defined from this. The techniques are standard, and the main results are direct consequences of the above properties. We suppose now that an antipode has been fixed until the end of this section. We will therefore omit the subscript when writing the orthogonality.

\begin{defi}
A \emph{type} of support $V$ is a set $\cond{A}$ of proof-objects of support $V$ such that there exists a set $B$ of proof-objects with $\cond{A}=B^{\pol}$. Equivalently, a type is a set $\cond{A}$ such that $\cond{A}=\cond{A}^{\pol\pol}$.
\end{defi}

The following constructions on type can then be shown to define a model of \MLL as it was done in previous work \cite{seiller-goim}. For two types $\cond{A}$ and $\cond{B}$, we define:
\begin{eqnarray*}
\cond{A\otimes B} &=& \{ \de{a}\plug\de{b} \mid \de{a}\in\cond{A}, \de{b}\in\cond{B}\}^{\pol\pol}\\
\cond{A\multimap B} &=& \{ \de{f} \mid \forall \de{a}\in\cond{A}, \de{f\plug a}\in\cond{B}\}
\end{eqnarray*}
A model of Multiplicative-Additive Linear Logic (\MALL) can also be constructed by considering linear combinations of proof-objects \cite{seiller-goiadd}. Both these constructions are quite automatic and the results are mainly dependent on the two properties cited above: associativity of execution and the trefoil property (here expressed as the cocycle Eq. (\ref{cocycle})).

\section{Graphings and dynamical systems \label{section:dynsys}\label{sec:graphings}}

In this section, we review the more general setting of Interaction Graphs based on graphings \cite{seiller-goig}. We first explain how the notions introduced in the previous section generalise, pinpointing how out the general construction based on zeta functions naturally adapts here. We then explain how deterministic graphings correspond to partial dynamical systems.

We first recall briefly the notion of graphing. Interested readers can find more detailed presentations in the author's recent work on computational complexity \cite{seiller-goinda,seiller-PRAMsLB}. The definition is parametrised by an \emph{abstract model of computation}: a monoid action $\alpha: M\acton\Space{X}$ on the underlying space $\Space{X}$, i.e. $\alpha$ is a monoid homomorphism from $M$ into the group of endomorphisms of $\Space{X}$. 

\begin{defi}[Abstract model of computation]
An \emph{abstract model of computation} (\amc) is a monoid action $\alpha: M\acton\Space{X}$.
\end{defi}

\begin{example}
Turing machines give rise to a monoid action as follows. One considers the space of configurations $\mathbf{X}=\{\ast,0,1\}^{\mathbb{|Z|}}$ of $\mathbf{Z}$-indexed sequences of symbols $\ast,0,1$ that are almost always equal to $\ast$. Then the monoid is generated by five maps: $\mathtt{left}$, $\mathtt{right}$, $\mathtt{write}_0$, $\mathtt{write}_1$, $\mathtt{write}_\ast$ acting on $\mathbf{X}$ as expected. For instance moving the working head to the right can be represented as the map $\mathtt{right}: \mathbf{X}\rightarrow \mathbf{X}, (a_i)_{i\in\mathbf{Z}}\mapsto (a_{i+1})_{i\in\mathbf{Z}}$, and the map $\mathtt{write}_0$ acts as $(a_i)_{i\in\mathbf{Z}}\mapsto (\hat{a}_{i})_{i\in\mathbf{Z}}$ where $\hat{a}_i=0$ if $i=0$ and $\hat{a}_i=a_i$ otherwise. 
\end{example}

Given an \amc $\alpha: M\acton\Space{X}$, one then defines $\alpha$-graphings (or \emph{abstract programs}) through the notion of \emph{$\alpha$-graphing representative}. A graphing representative is a \emph{geometric realisation} of a graph: it is a collection of pairs $(S,m)$ (called edges) where $S$ is a subspace of $\Space{X}$ -- the source of the edge -- and $m$ is an element of the monoid -- a sequence of instructions. For instance, in the Turing machines example above, the instruction \enquote{if the head is reading a $0$ or a $1$, move to the right} is represented as an edge $(S,m)$ whose source is the subspace $S=\{(a_i)_{i\in\mathbf{Z}}\in \mathbf{X}\mid a_0\neq \ast\}$ and realised by the map $\mathtt{right}$. We will leave it to the reader to convince herself that any Turing machine can be represented in this way.

\begin{defi}
An $\alpha$-graphing representative $G$ (w.r.t. a monoid action $\alpha:M\acton\Space{X}$) is defined as a set of \emph{edges} $E^{G}$ and for each element $e\in E^{G}$ a pair $(S^{G}_{e},m^{G}_{e})$ of a subspace $S^{G}_{e}$ of $\Space{X}$ -- the \emph{source} of $e$ -- and an element $m^{G}_{e}\in M$ -- the \emph{realiser} of $e$.
Similarly, a \emph{weighted $\alpha$-graphing representative} $G$ is defined as a set of edges $E^G$ and an $E^G$-indexed family of triples $\{ (S^G_e,m^G_e,\omega^G_e) \mid e\in E^G\}$.
\end{defi}

In the following, we will identify non-weighted graphings with weighted graphings with constant weight equal to $1$. Also, while the notion is quite general, we will restrict our discussion to the case of $\Space{X}$ being a measured space.

\begin{rem}
We note that in the general setting, a graphing also possesses a space of control states $Q^{G}$. The notions of source and realisers are then adapted: the source is a subset of $\Space{X}\times Q^{G}$, and the realiser is an element of $M\times \mathfrak{S}_{Q^{G}}$ -- where $\mathfrak{S}_{Q^{G}}$ is the group of permutations on $Q^{G}$.
While this generalisation is important for defining the models, we will only introduce control states in the section on Markov processes. This makes the results in this section (and the next) easier to state.
\end{rem}

An $\alpha$-graphing is then defined as an equivalence class of graphing representatives w.r.t. some notion of \emph{refinement}. The intuition is that an $\alpha$-graphing represents an \emph{action} on the underlying space, which can be defined by different graphing representatives. The base example is that of a graphing representative $G$ with a single edge $e$ of source $S_e$ and realised by the monoid element $m_e$, and the graphing representative $H$ with two edges $e_1,e_2$ of respective sources $S_{e_1}$ and $S_{e_2}$ and realised by $m_{e_1}=m_{e_2}=m_e$. The graphing representatives $G$ and $H$ represent the same action on the underlying space $\Space{X}$ as long as\footnote{Since we supposed $\Space{X}$ is a measured space, equalities holds up to a null measure set, while those can be exact in other cases, e.g. topological spaces.} $S_e=_{a.e.} S_{e_1}\cup S_{e_2}$ and $S_{e_1}\cap S_{e_2}=_{a.e} \emptyset$. In fact, $H$ is more than equivalent to $G$, it is a \emph{refinement} of the latter. 

\begin{defi}[Refinement]
A graphing representative $F$ is a refinement of a graphing representative $G$,
denoted $F\leqslant G$, if there exists a partition\footnote{We allow the sets
  $E^{F}_{e}$ to be empty.} $(E^{F}_{e})_{e\in E^{G}}$ of $E^{F}$ such that:
$$
\forall e\in E^{G}, \forall f\in E^{F}_{e},~ m^{F}_{f}=m_{e}^{G},
$$
$$
\forall e\in E^{G}, \forall f\neq f'\in E^{F}_{e},~ \mu(S^{F}_{f} \mathbin{\triangle}
S^{F}_{f'}) = 0,
$$
$$
\forall e\in E^{G}, \mu\left(\left(\cup_{f\in E^{F}_{e}} S^{F}_{f} \right) \mathbin{\triangle}
S^{G}_{e}\right) = 0.
$$
\end{defi}

Two graphing representatives $F$, $G$ are then equivalent (have the same action on the underlying space) whenever there exists a common refinement $H$, i.e. such that $H\leqslant F$ and $H\leqslant G$. The fact that this defines an equivalence relation compatible with the essential operations on graphing representatives to define models of linear logic (execution and measurement), is shown in the author's first work on graphings \cite{seiller-goig}.

\begin{defi}
An \emph{$\alpha$-graphing} (or an \emph{abstract program in the \amc $\alpha$}) is an equivalence class of $\alpha$-graphing representatives w.r.t. the equivalence relation generated by refinements: $F\sim G$ if and only if there exists $H$ with $H\leqslant F$ and $H\leqslant G$.
\end{defi}

We refer the interested reader to earlier papers \cite{seiller-goig,seiller-goif} for the definitions of execution and measurement of graphings \cite{seiller-goig}. We will write $\model{\Omega}{\alpha}$ for the obtained realisability model, where $\Omega$ is the monoid of weights (as already mentioned, we will only consider the case $\Omega=\complexN$ in this paper) and $\alpha$ the \amc. By extension, the notation $\model{\Omega}{\alpha}$ also denotes the set of all $\Omega$-weighted $\alpha$-graphings. We will now show how graphings relate to well-established notions in mathematics.

\subsection{Dynamical Systems}

Measured dynamical systems are a well-studied field of mathematics and apply to a range of physical and biological problems. We will not redefine basic notions of measure theory; those can be found in standard textbooks \cite{CohnMeasure}.

\begin{defi}
A \emph{measured dynamical system} is a pair $(\measure{X},f)$ of a measured space $\measure{X}$ and a measurable map $f:\measure{X}\rightarrow \measure{X}$. A \emph{partial measured dynamical system} is a triple $(\measure{X},D,f)$ where $\measure{X}$ is a measured space, $D\subseteq \measure{X}$ a subspace -- the domain --, and $f: D\rightarrow \mathbf{X}$ is a measurable map.
\end{defi}

The measured space $\measure{X}=(X,\mathcal{B},\mu)$ represents the set of states of the system under consideration, while the map $f$ describes the dynamics, i.e. the time-evolution of the system, based on the assumption that those do not vary with time (e.g. they are consequences of physical laws which are supposed not to change over time). It is then the iterated maps $f^{i}$ (and \emph{orbits} $\{ (f^i(x))_{i} \mid x\in\Space{X} \}$) that are of interest as they describe how the system will evolve.

Dynamical systems represent deterministic systems, such as those described by classical mechanics. If one wants to describe non-deterministic behaviour, one is led to consider several partial maps. The resulting object coincides with the notion of \emph{graphing} without weights. Describing probabilistic behaviour can be done by considering several partial maps assigned with probabilities; the resulting object is then a graphing with weights in $[0,1]$. While we will consider the latter case in the next section (where we will show that they correspond to a subclass of subprobabilistic kernels), we now focus on the deterministic case.

\begin{defi}
An $\alpha$-graphing representative $G=\{S^{G}_{e},\phi^{G}_{e},\omega^{G}_{e}~|~e\in E^{G}\}$ is \emph{deterministic} if $\forall e\in E^G, ~\omega_e^G=1$ and the following holds:
\[ \mu\left(\left\{x\in \measure{X}~|~ \exists e,f\in E^{G}, e\neq f\text{ and }x\in S_{e}^{G}\cap S_{f}^{G}\right\}\right)=0. \]
We note that this also defines the notion of \emph{deterministic graphing}: if $F$ is an $\alpha$-graphing representative equivalent to $G$ and $G$ is deterministic, then $F$ is deterministic.
\end{defi}

\begin{thm}\label{detdynsys}
There is a one-to-one correspondence between deterministic graphings and partial non-singular measurable-preserving dynamical systems (up to a.e. equality).
\end{thm}

\begin{proof}
Clearly, a partial dynamical system $(\measure{X},V,\Phi)$ where $\Phi$ is a \nsmp map defines a graphing of support $V$ with a single edge realised by $\Phi$. 

Now, let us explain how a deterministic graphing $G$ defines a partial non-singular measurable-preserving dynamical system. Since $G$ it is deterministic, we can consider a representative $\bar{G}$ of $G$ such that the set 
\[\left\{x\in \measure{X}~|~ \exists e,f\in E^{G}, e\neq f\text{ and }x\in S_{e}^{G}\cap S_{f}^{G}\right\}\]
 is the empty set. Then, one defines the partial dynamical system $(\measure{X},\cup_{e\in E^{\bar{G}}} S_{e}^{\bar{G}},\Phi)$, where:
\[\Phi(x)=\left\{
	\begin{array}{l}
	\phi_{e}^{\bar{G}}(x)\textrm{ if }x\in S_{e}^{\bar{G}}\\
	0\textrm{ otherwise}
	\end{array}
	\right.
	\]
Moreover the map $\Phi$ is \nsmp as a (disjoint) union of partial \nsmp maps. 

To end the proof, we need to show that the choice of representative in the previous construction is irrelevant. We prove this by showing that $G$ is equivalent to the graphing $H$ induced by $(\measure{X},\cup_{e\in E^{\bar{G}}} S_{e}^{\bar{G}},\Phi)$. But this is obvious, as $\bar{G}$ is a refinement of $H$, and $G$ and $\bar{G}$ are equivalent. This is sufficient because of the following claim: if $G$ and $G'$ are equivalent, then the induced partial dynamical systems are a.e. equal.
\end{proof}

More precisely, deterministic $\alpha$-graphings are in one-to-one correspondence with partial measured dynamical systems $(\Space{X},D,f)$ such that the graph of $f$ is included in the measurable preorder\footnote{If $\alpha$ is a group action by measure-preserving transformations, $\mathcal{P}(\alpha)$ is a \emph{Borel equivalence relation}, which can be used to construct von Neumann algebras with a distinguished maximal abelian subalgebra ({\sc masa}) \cite{FeldmanMoore1,FeldmanMoore2}. This is one of the intuitions behind the author's approach to complexity using graphings, since he established a correspondence between inclusions of {\sc masa}s in von Neumann algebras and the expressivity of the logical system arising from realisability techniques \cite{seiller-masas}.} $\mathcal{P}(\alpha)=\{ (x,y)\in\Space{X}\times\Space{X}\mid \exists m\in M, \alpha(m)(x)=y \}$.

\subsection{A submodel}

We now prove that the set of deterministic graphings is closed under the operation of execution, i.e. if $F,G$ are deterministic graphings, then their execution $F\plug G$ is again a deterministic graphing. This shows that the set of deterministic graphings defines a submodel $\dmodel{\Omega}{\alpha}$ of $\model{\Omega}{\alpha}$, i.e. it is a subset of graphings closed under execution and therefore defines a realisability model of linear logic, using the restriction of the measurement defined on $\model{\Omega}{\alpha}$.

\begin{lem}\label{lem:deterministicisclosed}
The execution of two deterministic graphings is a deterministic graphing.
\end{lem}

\begin{proof}
A \emph{deterministic graphing} $F$ satisfies that for every edges $e,f\in E^{F}$, $S^{F}_{e}\cap S^{F}_{f}$ is of null measure. Suppose that the graphing $F\plug G$ is not deterministic. Then there exists a Borel $B$ of non-zero measure and two edges $e,f\in E^{F\plug G}$ such that $B\subset S^{F\plug G}_{e}\cap S^{F\plug G}_{f}$. The edges $e,f$ correspond to paths $\pi_{e}$ and $\pi_{f}$ alternating between $F$ and $G$. It is clear that the first step of these paths belong to the same graphing, say $F$ without loss of generality, because the Borel set $B$ did not belong to the \emph{cut}. Thus $\pi_{e}$ and $\pi_{f}$ can be written $\pi_{e}=f_{0}\pi^{1}_{e}$ and $\pi_{f}=f_{0}\pi^{1}_{f}$ since $F$ is deterministic. Thus the domains of the paths $\pi^{1}_{e}$ and $\pi_{f}^{1}$ coincide on the Borel set $\phi_{f_{0}}^{F}(B)$ which is of non-zero measure since all maps considered are non-singular. One can then continue the reasoning up to the end of one of the paths and show that they are equal up to this point. Now, if one of the paths ends before the other we have a contradiction because it would mean that the Borel set under consideration would be at the same time inside and outside the cut, which is not possible. So both paths have the same length and are therefore equal. This shows that $F\plug G$ is deterministic since we have shown that if the domain of two paths alternating between $F$ and $G$ coincide on a non-zero measure Borel set, the two paths are equal (hence they correspond to the same edge in $F\plug G$).
\end{proof}

One can then check that the interpretations of proofs by graphings in earlier papers \cite{seiller-goig,seiller-goie,seiller-goif} are all deterministic. This gives us the following theorem as a corollary of the previous lemma.

\begin{thm}[Deterministic model]\label{thm:detmodel}
Let $\Omega$ be a monoid and $\alpha$ a monoid action. The set of $\Omega$-weighted \emph{deterministic} $\alpha$-graphings yields a model, denoted by $\dmodel{\Omega}{\alpha}$, of \MALL.
\end{thm}

This thus defines a realisability model $\dmodel{\Omega}{\alpha}$ of linear logic based on the set of all partial measured dynamical systems whose graph is included in $\mathcal{P}(\alpha)$, based on \autoref{detdynsys} and \autoref{thm:detmodel}. This is constructed using the measurement defined in earlier work \cite{seiller-goig} which, similarly to the graph case, we will now show is related to a standard notion of zeta function.

\subsection{Zeta Functions for dynamical systems}

The Ruelle zeta function \cite{Ruellezeta} is defined from a function $f: M\rightarrow M$, where $M$ is a manifold, and a matrix-valued  function $\phi: M\rightarrow \vn{M}_{k}$ (here $\vn{M}_{k}$ is the algebra of $k\times k$ matrices over the field of complex numbers). We write $\Fix{g}$ for the set of fixed points of $g$. Then the Ruelle zeta function is defined as (we suppose that $\Fix{f^{k}}$ is finite for all $k$):
$$ \zeta_{f,\phi}(z) = \exp\left(\sum_{m\geqslant 1}\frac{z^{m}}{m}\sum_{x\in\Fix{f^{m}}}\tr\left(\prod_{i=0}^{m-1}\phi(f^{i}(x))\right)\right). $$
For $k=1$ and $\phi=1$ the constant function equal to $1$, this is the Artin-Mazur \cite{artinmazur} zeta function:
$$ \zeta_{f,1}(z) = \exp\left(\sum_{m\geqslant 1}\frac{z^{m}}{m}\card{\Fix{f^{m}}}\right). $$

Since we work with measured spaces, we consider the following measured variant of Ruelle's zeta function (defined for measure-preserving maps\footnote{Based on the result of \autoref{proposition:zetagraphingmeas}, a definition for general \nsmp maps could be obtained using the method used by the author \cite{seiller-goig} to define a generalised measurement between graphings. However, we considered this to be out of the scope of this work.}). Suppose we work with a measured space $(M,\mathcal{B},\mu)$ and that $\Fix{f^{m}}$ is of finite measure:
$$ \zeta_{f,\Phi}(z) = \exp\left(\sum_{m\geqslant 1}\frac{z^{m}}{m}\int_{\Fix{f^{m}}}\tr\left(\prod_{i=0}^{m-1}\Phi(f^{i}(x))\right)d\mu(x)\right) $$
For $d=1$ and $\phi=1$, this becomes:
$$ \zeta_{f,1}(z) = \exp\left(\sum_{m\geqslant 1}\int_{\Fix{f^{m}}} \frac{z^{m}}{m}\right) $$
which we relate to the measurement on graphings defined in earlier work \cite{seiller-goig}.

\begin{prop}\label{proposition:zetagraphingmeas}
Given \nsmp partial dynamical systems $f,g:\mathbf{X}\rightarrow\mathbf{X}$, for all constant $c$ we have:
\[ \meas[\lambda x.c]{f,g}=\log(\zeta_{g\circ f,1}(c)), \]
with $\meas{\_,\_}$ the standard measurement on graphings \cite{seiller-goig}.
\end{prop}

\begin{proof}
On one hand, we have 
\[-\log(\zeta_{g\circ f,1}(1))=\int_{\Fix{(g\circ f)^{m}}} \frac{1}{m} \sum_{m\geqslant 1}\mu(\Fix{(g\circ f)^{m}}). \]

On the other hand, the measurement $\meas{f,g}$ defined on general graphings \cite[Definitions 37 and 57]{seiller-goig} is given by the formula
\[
\sum_{\pi=e_0e_1\dots e_n \in\PrimeCPath{f,g}} ~\sum_{j=0}^{n}\int_{\supp{\pi}} \sum_{k=0}^{\rho_{\phi_{\pi}}(x)-1}\frac{m(\omega(\pi)^{\rho_{\phi_{\pi}}(\phi_{\pi}^{k}(x))})}{(n+1)\rho_{\phi_\pi}(x)\rho_{\phi_{\pi}}(\phi_{\pi}^{k}(x))}d(\phi_{e_{n}}\circ\phi_{e_{n-1}}\circ\dots\circ\phi_{e_{j}})_{\ast}\lambda(x),
\]
where:
\begin{itemize}
\item $\rho_{\phi}$ is a measurable map associating to each point the length of the orbit it belongs to \cite[Corollary 45]{seiller-goim},
\item $\PrimeCPath{f,g}$ denotes the set of prime closed paths alternating between $f$ and $g$,  
\item and generally $h_\ast \mu$ denotes the pullback measure of $\mu$ along $h$.
\end{itemize}

As established by the author, this expression simplifies in the measure-preserving case \cite[Proposition 52]{seiller-goim}, and can be expressed as
\begin{equation*}
\meas{f,g}=\sum_{\pi=e_{0}\dots e_{n} \in\PrimeCPath{f,g}}\int_{\supp{\pi}} \frac{m(\omega(\pi)^{\rho_{\phi_{\pi}}(x)})}{\rho_{\phi_{\pi}}(x)}
\end{equation*} 

Now, we can split this expression by considering the partition of $\supp{\pi}$ given by the preimage of  $\rho_{\phi}$. Specifically, this partitions $\supp{\pi}$ into (measurable) subsets $S^\pi_i=\rho^{-1}_{\phi}(\supp{\pi})$ containing the points $x\in \supp{\pi}$ such that the orbit of $x$ is of length $i$.

As the value of $\rho_{\phi}$ is constant on these sets, this gives:
\begin{align*}
\meas{f,g} &= \sum_{\pi=e_{0}\dots e_{n} \in\PrimeCPath{f,g}}~\sum_{i=0}^{\infty}\int_{S^\pi_i} \frac{m(\omega(\pi)^{i})}{i}\nonumber
\end{align*}
Now, we are considering the case where $m(x)=z$, and we know all weights in the graphing are equal to $1$. Hence:
\begin{align*}
\meas{f,g} 
			&= \sum_{\pi \in\PrimeCPath{f,g}}~\sum_{i=0}^{\infty}\int_{S^\pi_i} \frac{z}{i}.\nonumber
\end{align*}

On the other hand, we have that, writing $\altpath{f,g}_m$ for the set of all alternating cycles between $f$ and $g$ of length $m$:
\begin{align*}
\log(\zeta_{g\circ f,1}(z))	&=\sum_{m\geq 1} \int_{\mathrm{Fix}((g\circ f)^m)} \frac{z}{m}\\
					&=\sum_{m\geq 1} ~\sum_{\pi\in\altpath{F,G}_m}\int_{S^\pi_m} \frac{z}{m}
\end{align*}
since each fixpoint belongs to exactly one alternating cycle of length $m$ between $f$ and $g$ (because the graphings are deterministic).

Now each alternating cycle of length $m$ between $f$ and $g$ can be written uniquely as a product of alternating prime cycles, we deduce (this is essentially Proposition 60 in the author's first paper on Interaction Graphs \cite{seiller-goim}):
\begin{align*}
\log(\zeta_{g\circ f,1}(z))	
				&=\sum_{m\geq 1} ~\sum_{\pi\in\PrimeCPath{f,g}}\int_{S^\pi_m} \frac{z}{m}\\
				&=\sum_{\pi\in\PrimeCPath{f,g}}~\sum_{m\geq 1} \int_{S^\pi_m} \frac{z}{m}\\
				&=\meas{f,g}
\end{align*}
This is the equality we wanted to prove.
\end{proof}

This shows that the author's realisability (sub)models of deterministic graphings, or equivalently of partial measured dynamical systems, can be constructed using zeta functions to define the orthogonality, similarly to the restricted graph setting we considered in the previous section.

\section{Probabilities and kernels \label{section:prob}}

As mentioned above, one could also consider a notion of \emph{probabilistic graphings} to represent probabilistic processes. Recall that we consider here $\Omega=\complexN$ so it makes sense to talk about weights taken in the interval $[0,1]$. We will show how this notion is closed under composition, and hence defines a \emph{sub-probabilistic model}, and how the corresponding objects capture specific subprobabilistic kernels.

\subsection{A probabilistic model}

\begin{defi}
A graphing $G=\{S^{G}_{e},\phi^{G}_{e},\omega^{G}_{e}~|~e\in E^{G}\}$ is \emph{sub-probabilistic} if the following holds:
\[ \mu\left(\left\{x\in \measure{X}~|~ \sum_{e\in E^{G}, x\in S^{G}_{e}}\omega^{G}_{e}>1\right\}\right)=0 \]
\end{defi}

It turns out that this notion of graphing also behaves well under composition, i.e. there exists a \emph{sub-probabilistic} submodel of $\model{\Omega}{\alpha}$, namely the model of \emph{sub-probabilistic graphings}. As explained below in the more general case of Markov processes (\autoref{whysub}), probabilistic graphings are \emph{not} closed under composition.

\begin{thm}\label{thm:probabilisticisclosed}
The execution of two sub-probabilistic graphings is a sub-probabilistic graphing.
\end{thm}

\begin{proof}
If the weights of edges in $F$ and $G$ are elements of $[0,1]$, then it is clear that the weights of edges in $F\plug G$ are also elements of $[0,1]$. We therefore only need to check that the second condition is preserved.

Let us denote by $\outset{F\plug G}$ the set of $x\in X$ which are sources of paths whose added weight is greater than $1$, and by $\outset{F\cup G}$ the set of $x\in X$ which are sources of edges (either in $F$ or $G$) whose added weight is greater than $1$. First, we notice that if $x\in\outset{F\plug G}$ then either $x\in\outset{F\cup G}$, or $x$ is mapped, through at least one edge, to an element $y$ which is itself in $\outset{F\cup G}$. To prove this statement, let us write $\outpaths{x}$ (resp. $\outedges{x}$) for the set of paths in $F\plug G$ (resp. edges in $F$ or $G$) whose source contain $x$. We know the sum of all the weights of these paths is greater than $1$, i.e. $\sum_{\pi\in\outpaths{x}}\omega(\pi)>1$. But this sum can be rearranged by ordering paths depending on their initial edge, i.e. 
\[ \sum_{\pi\in\outpaths{x}}\omega(\pi)=\sum_{e\in\outedges{x}}~\sum_{\pi=e\rho\in\outpaths{x}^{e}}\omega(\pi),\] 
where $\outpaths{x}^{e}$ denotes the paths whose first edge is $e$. Now, since the weight of $e$ appears in all $\omega(e\rho)=\omega(e)\omega(\rho)$, we can factorize and obtain the following inequality:
\[ \sum_{e\in\outedges{x}}\omega(e)\left(\sum_{\pi=e\rho\in\outpaths{x}^{e}}\omega(\rho)\right)>1 \] 
Since the sum $\sum_{e\in\outedges{x}}\omega(e)$ is not greater than $1$, we deduce that there exists at least one $e\in\outedges{x}$ such that $\sum_{\pi=e\rho\in\outpaths{x}^{e}}\omega(\rho)>1$. However, this means that $\phi_{e}(x)$ is an element of $\outset{F\plug G}$.

Now, we must note that $x$ is not an element of a closed path. This is clear from the fact that $x$ lies in the carrier of $F\plug G$.

Then, an induction shows that $x$ is an element of $\outset{F\plug G}$ if and only if there is a (finite, possibly empty) path from $x$ to an element of $\outset{F\cup G}$, i.e. $\outset{F\plug G}$ is at most a countable union of images of the set $\outset{F\cup G}$. But since all maps considered are non-singular, these images of $\outset{F\cup G}$ are negligible subsets since $\outset{F\cup G}$ is itself negligible. This ends the proof as a countable union of copies of negligible sets is negligible (by countable additivity), hence $\outset{F\plug G}$ is negligible.
\end{proof}

As a corollary, we get an equivalent of \autoref{thm:detmodel}.

\begin{thm}[Probabilistic model]
Let $\alpha:M\acton \mathbf{X}$ be a monoid action. The set of $\Omega$-weighted \emph{probabilistic} $\alpha$-graphings yields a model, denoted by $\pmodel{\Omega}{\alpha}$, of Multiplicative-Additive Linear Logic.
\end{thm}

We now explain how these models can be understood as realisability models over a subclass of (sub-)Markov processes.

\subsection{Discrete-image sub-Markov processes}

We are now considering probabilistic systems. More specifically, we consider systems for which evolution is still time-independent, but which obey the principle of probabilistic choices: given a state, it may produce different outputs but these different choices are provided with a probability distribution. The notion of a dynamical system, i.e. a map from a measured space to itself, is then no longer the right object to formalise this idea. In fact, a probabilistic time evolution does not act on the states of the system but rather on the set of probability distributions on this set of states.

\begin{defi}
Let $\measure{X}$ be a measured space. We denote $\distribs{X}$ the set of sub-probability distributions over $\measure{X}$, i.e. the set of sub-probability measures on $\measure{X}$.
\end{defi} 

Now, a deterministic system also acts on the set of probability measures by post-composition. If $(\measure{X},f)$ is a measured dynamical system, then given a (sub-)probability distribution (otherwise called a \emph{random variable}) $p:\measure{P}\rightarrow\measure{X}$, the map $f\circ p$ is itself a (sub-) probability distribution. This action of deterministic graphings (equivalently, dynamical systems) on the set of (sub-)probability distributions $\distribs{X}$ can be naturally extended to an action of sub-probabilistic graphings on $\distribs{X}$. In fact, we show that sub-probabilistic graphings define sub-Markov kernels. We recall that sub-probability distributions on $\measure{X}$ are Markov kernels from the one-point space $\{\ast\}$ to $\measure{X}$, and the action of a sub-Markov kernel onto $\distribs{X}$ is defined as post-composition (using the composition of kernels) \cite{CategoryKernels}.

\begin{notations}
In this section and the following, we write measured spaces $\measure{X}$, $\measure{Y}$, etc. in boldface fonts. We will use the same letter in normal fonts, e.g. $X$, $Y$, etc. to denote the underlying set and the same letter in calligraphic fonts, e.g. $\mathcal{X}$, $\mathcal{Y}$, etc. to denote the associated $\sigma$-algebra. We do not assume generic notation for the measures and, should the need to talk about them arise, we would explicitly name them.
\end{notations}

\begin{defi}
Let $\measure{X}$, $\measure{Y}$ be measured spaces. A sub-Markov kernel on $\measure{X}\times\measure{Y}$ is a measurable map $\kappa: X\times \mathcal{Y}\rightarrow [0,1]$ such that $\forall x\in X$ and $\forall B\in \mathcal{Y}$, $\kappa(x,\_)$ is a subprobability measure on $X$ and $\kappa(\_,B)$ is a measurable function.
If $\kappa(x,\_)$ is a probability measure, $\kappa$ is a Markov kernel.
\end{defi}

\begin{defi}
A \emph{discrete-image kernel} is a sub-Markov kernel $\kappa$ on $\measure{X}\times \measure{Y}$ such that for all $x\in\measure{X}$, $\kappa(x,\_)$ is a discrete probability distribution.
\end{defi}

\begin{notations}
To simplify equations, we write $\dot{x}$ instead of the usual $dx$ (or $d\mu(x)$) in the equations. With this notation, the composition of the kernels $\kappa$ on $\measure{X}\times \measure{Y}$ and $\kappa'$ on $\measure{Y}\times \measure{Z}$ is computed as follows:
\[ \kappa'\circ\kappa(x,\dot{z}) = \int_{Y}\kappa(x,\dot{y})\kappa'(y,\dot{z}). \]
\end{notations}

\begin{thm}
There is a one-to-one correspondence between sub-probabilistic graphings on $\mathbf{X}$ and discrete-image sub-Markov kernels on $\measure{X}\times\measure{X}$.
\end{thm}

\begin{proof}
The fact that sub-probabilistic graphings define sub-Markov processes is quite easy. One defines from a graphing $G=\{S^{G}_{e},\phi^{G}_{e},\omega^{G}_{e}~|~e\in E^{G}\}$ the kernel:
\[ \kappa_G: X\times\mathcal{X}\rightarrow [0,1]; (x,Y)\mapsto \sum_{e\in E^G, x\in S^G_e, \phi^G_e(x)\in Y} \omega^{G}_e. \]
The fact that it is a discrete-image sub-Markov kernel is clear.

The converse, i.e. given a kernel $\kappa$, define a graphing $G_\kappa$ is more involved. The difficulty lies in the fact that one has to collect the pairs $(x,y)$ such that $\kappa(x,y)>0$ into a countable collection of measurable maps. The key ingredients to make this work are: the countability of $\{ Y\in \mathcal{X}\mid \kappa(x,Y)>0\}$ for all $x\in X$ (because $\kappa$ is supposed to be a discrete-image kernel), the possibility to approximate all real numbers by a (countable) sequence of rational numbers, the measurability of $\kappa(\_,B)$ for all $B\in\mathcal{X}$.
\end{proof}

As a consequence of the results in this section, the author's work \cite{seiller-goif} gives rise, when restricting to subprobabilistic graphings, to a realisability model of linear logic over discrete-image sub-Markov kernels. We now have the needed context to address the main question answered (positively) in this work: can one construct a realisability model of linear logic on the set of (unconstrained) sub-Markov processes?

\section{A sub-Markov processes cocycle\label{section:markovzeta}}

Based on the previous sections, we will now extend the realisability constructions to general Markov sub-processes. The need to consider sub-Markov kernels and not only Markov kernels is explained by technical reasons we illustrate below (\autoref{whysub}). In this section, we will assume all measurable spaces to be $\sigma$-finite.

\begin{notations}
In the following we write $\identity$ the \emph{identity kernel} on $\measure{X}\times\measure{X}$, i.e. the Dirac delta function $\identity(x,\dot{x})=\delta(x,\dot{x})$ s.t. $\int_{A}\identity(x,\dot{x})=1$ if $x\in A$ and $\int_{A}\identity(x,\dot{x})=0$ otherwise.
\end{notations}

We will now define the two key ingredients of the model: the execution and the zeta function. We will then proceed to prove the cocycle property which will ensure the realisability model obtained captures the linear logic discipline. 

\subsection{Execution and Zeta}

\begin{defi}[Iterated kernel]
Let $\kappa$ be a sub-Markov kernel on $\measure{X}\times \measure{Y}$. For $k>1$, we define the $k$-th iterated kernel:
\[ \kappa^{(k)}(x_0,\dot{x}_k) = \iint_{(x_1,\dots,x_{k-1})\in (\mathbf{X}\cap \mathbf{Y})^{k-1}} \prod_{i=0}^{k-1} \kappa(x_{i},\dot{x}_{i+1}). \]
By convention, $\kappa^{(1)}=\kappa$.
\end{defi}

\begin{defi}[Maximal paths -- Execution kernel]
Let $\kappa$ be a sub-Markov kernel on $\measure{X}\times \measure{Y}$. We define the \emph{execution kernel} of $\kappa$ as the map (in the formula, $x_{n+1}$ is used as a notation for $y$):
\[
\begin{array}{rcccl}
\tr(\kappa)&:&	X\backslash Y\times \mathcal{Y}\backslash \mathcal{X} &\rightarrow & [0,1]\\
		&&	(x,y) &\mapsto& \sum_{n\geqslant 1} \kappa^{(n)}(x,y).
\end{array}
\]
\end{defi}

\begin{rem}\label{whysub}
One could wonder why this is not defined on the whole space $X\times \mathcal{Y}$. The restriction is needed to define a sub-Markov kernel. This can be understood on a very simple Markov chain:
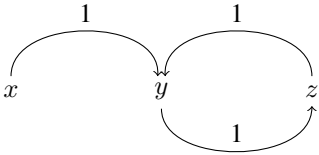
\begin{center}
\begin{tikzpicture}[y=0.7cm]
\node (x) at (0,0) {$x$};
\node (y) at (2,0) {$y$};
\node (z) at (4,0) {$z$};

\draw[->] (x) .. controls (0,1) and (2,1) .. (1.95,0.2) node [midway,above] {1};
\draw[->] (z) .. controls (4,1) and (2,1) .. (2.05,0.2) node [midway,above] {1};
\draw[->] (y) .. controls (2,-1) and (4,-1) .. (z) node [midway,above] {1};
\end{tikzpicture}
\end{center}
For this figure, the partial sums of $\kappa^{(i)}(x,y)$ is a diverging series. This example also shows why the resulting kernel could be a sub-Markov kernel even when $\kappa$ is a proper Markov kernel. 
\end{rem}

\begin{lem}\label{tracekernel}
If $\kappa$ is a sub-Markov kernel, $\tr(\kappa)$ is well-defined and a sub-Markov kernel.
\end{lem}

\begin{proof}
The gist of the proof is an induction to establish that for all integer $k$ and measurable subset $A$ such that $A\cap \mathbf{X}\cap \mathbf{Y}=\emptyset$, the expression $\int_{a\in A} \sum_{i=1}^{k} \kappa^{(i)}(x,\dot{a})$ is bounded by $1$. This is clear for $k=1$ from the assumption that $\kappa$ is a sub-Markov kernel. The following computation then establishes the induction (we write $x=y_0$ to simplify the equations):
\begin{equation*}
\int_{a\in A} \sum_{i=1}^{k+1} \kappa^{(i)}(y_0,\dot{a}) = \int_{a\in A} \kappa(y_0,\dot{a}) + \int_{a\in A} \sum_{i=0}^{k} \kappa^{(i+1)}(y_0,\dot{a}).
 \end{equation*}
 We now bound the second term as follows, using the induction hypothesis to establish that $\int_{a\in A}\sum_{i=1}^{k}\kappa^{(i)}(y_0,a)\leqslant 1$:
 \begin{align*}
 \lefteqn{\int_{a\in A} \sum_{i=0}^{k} \kappa^{(i+1)}(y_0,\dot{a})}\\
&= \int_{a\in A} \sum_{i=0}^{k} \int_{y_1}\dots\int_{y_i} \kappa(y_i,\dot{a})\prod_{j=0}^{i-1}\kappa(y_j,\dot{y}_{j+1})\\
 &=\int_{y_1}\int_{a\in A}\sum_{i=0}^{k} \int_{y_2}\dots\int_{y_i}  \kappa(y_i,\dot{a})\prod_{j=0}^{i-1}\kappa(y_j,\dot{y}_{j+1})\\
  &=  \int_{y_1}\kappa(y_0,\dot{y}_1)\int_{a\in A}\sum_{i=1}^{k} \int_{y_2}\dots\int_{i} \kappa(y_i,\dot{a})\prod_{j=0}^{i-1}\kappa(y_j,\dot{y}_{j+1})\\
  &=  \int_{y_1}\kappa(y_0,\dot{y}_1)\int_{a\in A}\sum_{i=1}^{k}\kappa^{(i)}(y_1,\dot{a})\\
  &\leqslant \int_{y_1}\kappa(y_0,\dot{y}_1).
\end{align*}
Coming back to the initial expression, we obtain it using the additivity of $\kappa$ (we recall that $A$ and $\mathbf{X}\cap \mathbf{Y}$ do not intersect):
\[ \int_{a\in A} \sum_{i=1}^{k+1} \kappa^{(i)}(y_0,\dot{a}) \leqslant \kappa(y_0,A)+\kappa(y_0,\mathbf{X}\cap \mathbf{Y}) \leqslant 1,\]
which is the required bound.
\end{proof}

Now, the execution kernel just defined is the main operation for defining the execution of sub-Markov kernels, as we will explain in section \ref{section:markovmodel}. We now define the second ingredient: the zeta function. For this, we first define a map which we call the \enquote{zeta kernel}.

\begin{defi}[Finite orbits -- Zeta kernel]
Let $\kappa$ be a sub-Markov kernel on $\measure{X}\times\measure{Y}$. The \emph{zeta kernel}, or kernel of finite orbits of $\kappa$, is a kernel on $\mathbf{X}\times\times\measure{Y}\naturalN$  (where $\naturalN$ denotes the set of natural numbers) defined as:
\[ \zeta_{\kappa}(x_0,\dot{x}_0,n)=\iint_{(x_1,\dots,x_{n-1})\in (\mathbf{X}\cap \mathbf{Y})^{n-1}} \prod_{i\in \mathbf{Z}\!/\!n\mathbf{Z}} \kappa(x_{i},\dot{x}_{i+1}).\]
\end{defi}

This expression computes the probability that a given point $x_0$ lies in an orbit of length $n$. It is a sub-Markov kernel for each fixed value of $n$, but the sum over $n\in\integerN$ is not. The reason is simple: if a point $x$ lies in a length $2$ orbit with probability $1$ (e.g. the point $y$ in the example Markov chain in \autoref{whysub}), then it lies in a length $2k$ orbit with probability $1$ as well. However, let us remark that the expression
\[ \int_{x\in X\cap Y}\zeta_{\kappa}(x,\dot{x},n) \]
plays the role of the set $\Fix{f^n}$ that appears in dynamical and graph zeta functions.

\begin{defi}[Zeta function]
Let $\kappa$ be a sub-Markov kernel on $\measure{X}\times \measure{Y}$. We define the Zeta function associated with $\kappa$ by:
\[ \zeta_{\kappa}(z): z\mapsto \exp\left(\sum_{n=1}^{\infty} \frac{z^n}{n}\int_{x\in \mathbf{X}\cap \mathbf{Y}} \zeta_{\kappa}(x,\dot{x},n)\right) \]
\end{defi}

\subsection{Execution and the Cocycle Property}

We will now define the execution of kernels and establish the two essential properties needed to construct a model of linear logic: associativity of execution, and the cocycle property satisfied by the zeta function.

\begin{defi}
Given two sub-Markov kernels $\kappa$ on $\mathbf{X}\times \mathbf{X'}$ and $\kappa'$ on $\mathbf{Y}\times \mathbf{Y'}$, we define their execution $\kappa\plug\kappa'$ as the kernel $\tr(\kappa\bullet\kappa')$ where:
\[
\kappa\bullet\kappa'= (\kappa'+\identity[\mathbf{X'}\backslash \mathbf{Y}])\circ (\kappa+\identity[\mathbf{Y}\backslash \mathbf{X'}])
\]
\end{defi}

Reader familiar with \emph{traced monoidal categories} \cite{tracedmonoidal,Hasegawatraced,catgoi} should not be surprised of this definition and the following properties\footnote{In fact, the execution kernel should define a trace in the categorical sense.}.

\begin{defi}
Three sub-Markov kernels $\kappa$ on $\mathbf{X}\times \mathbf{X'}$, $\kappa'$ on $\mathbf{Y}\times \mathbf{Y'}$, and $\kappa''$ on $\mathbf{Z}\times \mathbf{Z'}$ are said to be \emph{in general position}\footnote{The reader will realise the terminology is inspired from algebraic geometry, but no formal connections should be expected.} when the following condition is met:
\[
\mu(\mathbf{X'}\cap \mathbf{Y}\cap \mathbf{Z}) = \mu(\mathbf{Y'}\cap \mathbf{Z} \cap \mathbf{X}) = \mu(\mathbf{Z'}\cap \mathbf{X}\cap \mathbf{Y})=0,
\]
\[
\mu(\mathbf{X}\cap \mathbf{Y'}\cap \mathbf{Z'}) = \mu(\mathbf{Y}\cap \mathbf{Z'} \cap \mathbf{X'}) = \mu(\mathbf{Z}\cap \mathbf{X'}\cap \mathbf{Y'})=0.
\]
Note that if $\mathbf{X}=\mathbf{X'}$, $\mathbf{Y}=\mathbf{Y'}$ and $\mathbf{Z}=\mathbf{Z'}$, the condition becomes $\mu(\mathbf{X}\cap\mathbf{Y}\cap\mathbf{Z})=0$, which is the condition of application of the associativity of execution and of the trefoil property in the graph case.
\end{defi}

\begin{lem}\label{assocexec2}
Given three sub-Markov kernels $\kappa_0$ on $\mathbf{X}\times \mathbf{X'}$, $\kappa_1$ on $\mathbf{Y}\times \mathbf{Y'}$, and $\kappa_2$ on $\mathbf{Z}\times \mathbf{Z'}$ in general position:
\[
(\kappa_0\plug\kappa_1)\plug \kappa_2=\kappa_0\plug(\kappa_1\plug \kappa_2).
\]
\end{lem}

\begin{proof}
The fact that the Markov kernels are in general position allows us to write the composition in a traced monoidal category style (\autoref{fig21}).

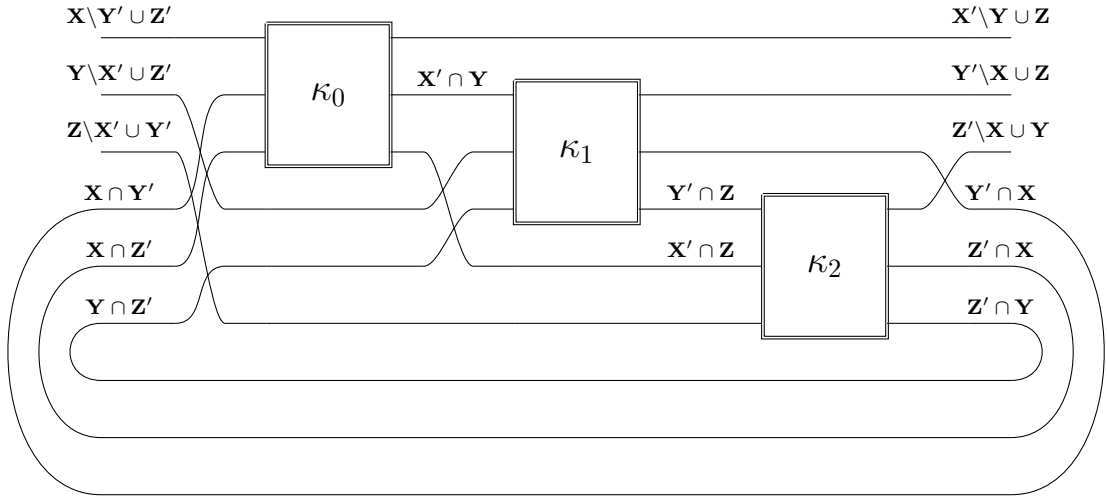
\begin{figure*}
\begin{center}
\begin{tikzpicture}[x=0.55cm,y=0.38cm]
	\node (X-Y'-Z') at (-0.3,0) {};
	\node (Y-X'-Z') at (-0.3,-2) {};
	\node (Z-X'-Y') at (-0.3,-4) {};
	
	\node (XcapY') at (0,-6) {};
	\node (XcapZ') at (0,-8) {};
	\node (YcapZ') at (0,-10) {};
	
	\draw[-] (-1,0) -- (0.8,0) node [near start,above] {\footnotesize{$\mathbf{X}\backslash\mathbf{Y'}\cup\mathbf{Z'}$}};
	\draw[-] (0.8,0) -- (3,0) {};
	\draw[-] (-1,-6) -- (0.8,-6) node [near start,above] {\footnotesize{$\mathbf{X}\cap\mathbf{Y'}$}};
	\draw[-] (0.8,-6) .. controls (1.6,-6) and (1.2,-2) .. (2,-2) -- (3,-2) {};
	\draw[-] (-1,-8) -- (0.8,-8) node [near start,above] {\footnotesize{$\mathbf{X}\cap\mathbf{Z'}$}};
	\draw[-] (0.8,-8) .. controls (1.6,-8) and (1.2,-4) .. (2,-4) -- (3,-4) {};
	\draw[-] (-1,-10) -- (0.8,-10) node [near start,above] {\footnotesize{$\mathbf{Y}\cap\mathbf{Z'}$}};
	\draw[-] (0.8,-10) .. controls (1.6,-10) and (1.2,-8) .. (2,-8) -- (3,-8) {};
	\draw[-] (-1,-4) -- (0.8,-4) node [near start,above] {\footnotesize{$\mathbf{Z}\backslash\mathbf{X'}\cup\mathbf{Y'}$}};
	\draw[-] (0.8,-4) .. controls (1.2,-4) and (1.6,-10) .. (2,-10) -- (3,-10) {};
	\draw[-] (-1,-2) -- (0.8,-2) node [near start,above] {\footnotesize{$\mathbf{Y}\backslash\mathbf{X'}\cup\mathbf{Z'}$}};
	\draw[-] (0.8,-2) .. controls (1.2,-2) and (1.6,-6) .. (2,-6) -- (3,-6) {};
	
	\draw[-,double] (3,0.5) -- (6,0.5) -- (6,-4.5) -- (3,-4.5) -- (3,0.5) {};
	\node (kappa) at (4.5,-2) {\Large{$\kappa_0$}};
	
	\draw[-] (6,0) -- (9,0) {};
	\draw[-] (6,-2) -- (9,-2) node [midway,above] {\footnotesize{$\mathbf{X'}\cap\mathbf{Y}$}};
	\draw[-] (6,-4) -- (6.8,-4) .. controls (7.2,-4) and (7.6,-8) .. (8,-8) -- (9,-8) {};
	\draw[-] (3,-6) -- (6.8,-6) .. controls (7.2,-6) and (7.6,-4) .. (8,-4) -- (9,-4) {};
	\draw[-] (3,-8) -- (6.8,-8) .. controls (7.2,-8) and (7.6,-6) .. (8,-6) -- (9,-6) {};
	\draw[-] (3,-10) -- (9,-10) {};

	\draw[-,double] (9,-1.5) -- (12,-1.5) -- (12,-6.5) -- (9,-6.5) -- (9,-1.5) {};
	\node (kappa') at (10.5,-4) {\Large{$\kappa_1$}};
	
	\draw[-] (9,0) -- (15,0) {};
	\draw[-] (12,-2) -- (15,-2) {};
	\draw[-] (12,-4) -- (15,-4) {};
	\draw[-] (12,-6) -- (15,-6) node [midway, above] {\footnotesize{$\mathbf{Y'}\cap\mathbf{Z}$}};
	\draw[-] (9,-8) -- (15,-8) node [near end, above] {\footnotesize{$\mathbf{X'}\cap\mathbf{Z}$}};
	\draw[-] (9,-10) -- (15,-10) {};

	\draw[-,double] (15,-5.5) -- (18,-5.5) -- (18,-10.5) -- (15,-10.5) -- (15,-5.5) {};
	\node (kappa'') at (16.5,-8) {\Large{$\kappa_2$}};
	
	\draw[-] (15,0) -- (20,0) {};
	\draw[-] (20,0) -- (21,0) node [near end,above] {\footnotesize{$\mathbf{X'}\backslash\mathbf{Y}\cup\mathbf{Z}$}};
	\draw[-] (15,-2) -- (20,-2) {};
	\draw[-] (20,-2) -- (21,-2) node [near end,above] {\footnotesize{$\mathbf{Y'}\backslash\mathbf{X}\cup\mathbf{Z}$}};
	\draw[-] (15,-4) -- (18.8,-4) .. controls (19.2,-4) and (19.6,-6) .. (20,-6) {};
	\draw[-] (20,-6) -- (21,-6) node [near end,above] {\footnotesize{$\mathbf{Y'}\cap\mathbf{X}$}};
	\draw[-] (18,-6) -- (18.8,-6) .. controls (19.2,-6) and (19.6,-4) .. (20,-4) {};
	\draw[-] (20,-4) -- (21,-4) node [near end,above] {\footnotesize{$\mathbf{Z'}\backslash\mathbf{X}\cup\mathbf{Y}$}};
	\draw[-] (18,-8) -- (20,-8) {};
	\draw[-] (20,-8) -- (21,-8) node [near end,above] {\footnotesize{$\mathbf{Z'}\cap\mathbf{X}$}};
	\draw[-] (18,-10) -- (20,-10) {};
	\draw[-] (20,-10) -- (21,-10) node [near end,above] {\footnotesize{$\mathbf{Z'}\cap\mathbf{Y}$}};

	\draw[-] (21,-12) -- (-1,-12) {};
	\draw[-] (21,-10) .. controls (22,-10) and (22,-12) .. (21,-12) {};
	\draw[-] (-1,-10) .. controls (-2,-10) and (-2,-12) .. (-1,-12) {};
	\draw[-] (21,-14) -- (-1,-14) {};
	\draw[-] (21,-8) .. controls (23,-8) and (23,-14) .. (21,-14) {};
	\draw[-] (-1,-8) .. controls (-3,-8) and (-3,-14) .. (-1,-14) {};
	\draw[-] (21,-16) -- (-1,-16) {};
	\draw[-] (21,-6) .. controls (24,-6) and (24,-16) .. (21,-16) {};
	\draw[-] (-1,-6) .. controls (-4,-6) and (-4,-16) .. (-1,-16) {};

\end{tikzpicture}
\end{center}\caption{Proof of \autoref{assocexec2}, first figure}\label{fig21}
\end{figure*}

The above theorem then states that the feedbacks commute with the composition. More precisely, it states that \autoref{fig22} computes the same kernel as the one below which represents the left-hand side of the equation. 

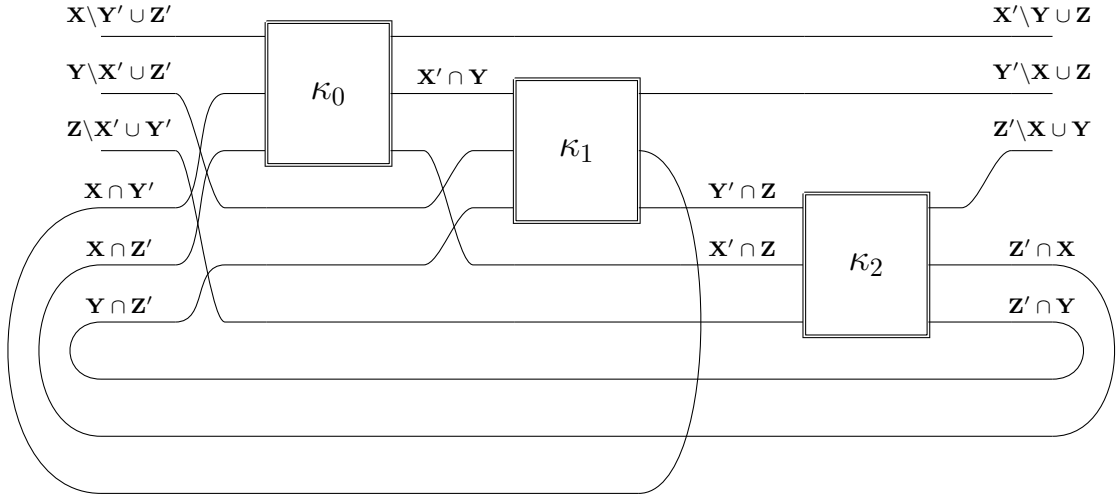
\begin{figure*}
\begin{center}
\begin{tikzpicture}[x=0.55cm,y=0.38cm]
	\node (X-Y'-Z') at (-0.3,0) {};
	\node (Y-X'-Z') at (-0.3,-2) {};
	\node (Z-X'-Y') at (-0.3,-4) {};
	
	\node (XcapY') at (0,-6) {};
	\node (XcapZ') at (0,-8) {};
	\node (YcapZ') at (0,-10) {};
	
	\draw[-] (-1,0) -- (0.8,0) node [near start,above] {\footnotesize{$\mathbf{X}\backslash\mathbf{Y'}\cup\mathbf{Z'}$}};
	\draw[-] (0.8,0) -- (3,0) {};
	\draw[-] (-1,-6) -- (0.8,-6) node [near start,above] {\footnotesize{$\mathbf{X}\cap\mathbf{Y'}$}};
	\draw[-] (0.8,-6) .. controls (1.6,-6) and (1.2,-2) .. (2,-2) -- (3,-2) {};
	\draw[-] (-1,-8) -- (0.8,-8) node [near start,above] {\footnotesize{$\mathbf{X}\cap\mathbf{Z'}$}};
	\draw[-] (0.8,-8) .. controls (1.6,-8) and (1.2,-4) .. (2,-4) -- (3,-4) {};
	\draw[-] (-1,-10) -- (0.8,-10) node [near start,above] {\footnotesize{$\mathbf{Y}\cap\mathbf{Z'}$}};
	\draw[-] (0.8,-10) .. controls (1.6,-10) and (1.2,-8) .. (2,-8) -- (3,-8) {};
	\draw[-] (-1,-4) -- (0.8,-4) node [near start,above] {\footnotesize{$\mathbf{Z}\backslash\mathbf{X'}\cup\mathbf{Y'}$}};
	\draw[-] (0.8,-4) .. controls (1.2,-4) and (1.6,-10) .. (2,-10) -- (3,-10) {};
	\draw[-] (-1,-2) -- (0.8,-2) node [near start,above] {\footnotesize{$\mathbf{Y}\backslash\mathbf{X'}\cup\mathbf{Z'}$}};
	\draw[-] (0.8,-2) .. controls (1.2,-2) and (1.6,-6) .. (2,-6) -- (3,-6) {};
	
	\draw[-,double] (3,0.5) -- (6,0.5) -- (6,-4.5) -- (3,-4.5) -- (3,0.5) {};
	\node (kappa) at (4.5,-2) {\Large{$\kappa_0$}};
	
	\draw[-] (6,0) -- (9,0) {};
	\draw[-] (6,-2) -- (9,-2) node [midway,above] {\footnotesize{$\mathbf{X'}\cap\mathbf{Y}$}};
	\draw[-] (6,-4) -- (6.8,-4) .. controls (7.2,-4) and (7.6,-8) .. (8,-8) -- (9,-8) {};
	\draw[-] (3,-6) -- (6.8,-6) .. controls (7.2,-6) and (7.6,-4) .. (8,-4) -- (9,-4) {};
	\draw[-] (3,-8) -- (6.8,-8) .. controls (7.2,-8) and (7.6,-6) .. (8,-6) -- (9,-6) {};
	\draw[-] (3,-10) -- (9,-10) {};

	\draw[-,double] (9,-1.5) -- (12,-1.5) -- (12,-6.5) -- (9,-6.5) -- (9,-1.5) {};
	\node (kappa') at (10.5,-4) {\Large{$\kappa_1$}};
	
	\draw[-] (9,0) -- (16,0) {};
	\draw[-] (12,-2) -- (16,-2) {};
	\draw[-] (12,-6) -- (13,-6) {};
	\draw[-] (13,-6) -- (16,-6) node [above, midway] {\footnotesize{$\mathbf{Y'}\cap\mathbf{Z}$}};
	\draw[-] (9,-8) -- (13,-8) {};
	\draw[-] (13,-8) -- (16,-8) node [above, midway] {\footnotesize{$\mathbf{X'}\cap\mathbf{Z}$}};
	\draw[-] (9,-10) -- (16,-10) {};

	\draw[-,double] (16,-5.5) -- (19,-5.5) -- (19,-10.5) -- (16,-10.5) -- (16,-5.5) {};
	\node (kappa'') at (17.5,-8) {\Large{$\kappa_2$}};
	
	\draw[-] (16,0) -- (21,0) {};
	\draw[-] (21,0) -- (22,0) node [near end,above] {\footnotesize{$\mathbf{X'}\backslash\mathbf{Y}\cup\mathbf{Z}$}};
	\draw[-] (16,-2) -- (21,-2) {};
	\draw[-] (21,-2) -- (22,-2) node [near end,above] {\footnotesize{$\mathbf{Y'}\backslash\mathbf{X}\cup\mathbf{Z}$}};
	\draw[-] (19,-6) -- (19.8,-6) .. controls (20.2,-6) and (20.6,-4) .. (21,-4) {};
	\draw[-] (21,-4) -- (22,-4) node [near end,above] {\footnotesize{$\mathbf{Z'}\backslash\mathbf{X}\cup\mathbf{Y}$}};
	\draw[-] (19,-8) -- (21,-8) {};
	\draw[-] (21,-8) -- (22,-8) node [near end,above] {\footnotesize{$\mathbf{Z'}\cap\mathbf{X}$}};
	\draw[-] (19,-10) -- (21,-10) {};
	\draw[-] (21,-10) -- (22,-10) node [near end,above] {\footnotesize{$\mathbf{Z'}\cap\mathbf{Y}$}};

	\draw[-] (22,-12) -- (-1,-12) {};
	\draw[-] (22,-10) .. controls (23,-10) and (23,-12) .. (22,-12) {};
	\draw[-] (-1,-10) .. controls (-2,-10) and (-2,-12) .. (-1,-12) {};
	\draw[-] (22,-14) -- (-1,-14) {};
	\draw[-] (22,-8) .. controls (24,-8) and (24,-14) .. (22,-14) {};
	\draw[-] (-1,-8) .. controls (-3,-8) and (-3,-14) .. (-1,-14) {};
	\draw[-] (12,-16) -- (-1,-16) {};
	\draw[-] (12,-4) .. controls (14,-4) and (14,-16) .. (12,-16) {};
	\draw[-] (-1,-6) .. controls (-4,-6) and (-4,-16) .. (-1,-16) {};

\end{tikzpicture}
\end{center}\caption{Proof of \autoref{assocexec2}, second figure}\label{fig22}
\end{figure*}

The fact that this is true is a consequence of the fact that kernels are in general position since the integrals are taken over disjoint domains. The underlying explanation is that the execution kernel $\kappa\plug\kappa'$ is computed by integrating over \emph{alternating paths} between $\kappa$ and $\kappa'$, i.e. it is computed as an integral over the sequences $x_1,x_2,\dots,x_k$ of the alternating product of $\kappa(x_{i},\dot{x}_{i+1})$ and $\kappa'(x_{i+1},\dot{x}_{i+2})$. These paths can be seen in the above figures. Taking the iterated composition $(\kappa_0\plug\kappa_1)\plug \kappa_2$ thus integrates over alternating paths between $\kappa_2$ and alternating paths between $\kappa_0$ and $\kappa_1$. Using the geometric identity relating alternating paths and cycles (Equation \ref{geometrictrefoil} on page \pageref{geometrictrefoil}, established in \cite{seiller-goiadd}), this is the same as integrating over all alternating paths between $\kappa_0$ and alternating paths between $\kappa_1$ and $\kappa_2$.
\end{proof}

\begin{lem}\label{commutexec}
Given two sub-Markov kernels $\kappa$ on $\mathbf{X}\times \mathbf{X}'$ and $\kappa'$ on $\mathbf{Y}\times \mathbf{Y}'$ such that $\mathbf{X}\cap\mathbf{Y}=\mathbf{X}'\cap\mathbf{Y}'=\emptyset$:
\[
\kappa\plug\kappa'=\kappa'\plug\kappa.
\]
\end{lem}

\begin{proof}
The assumption on the spaces implies that one can picture the composition as shown in Figure \ref{fig:commutexec}. \qedhere

\begin{figure*}
\begin{center}
\begin{tikzpicture}[x=0.55cm,y=0.38cm]
	\draw[-] (-1,4) -- (0.8,4) node [near start,above] {\footnotesize{$\mathbf{X}\backslash\mathbf{X}'\cup\mathbf{Y'}$}};
	\draw[-] (0.8,4) -- (1.8,4) .. controls (2,4) and (2.2,3.8) .. (2.2,3) ..controls (2.2,2.2) and (2.4,2) .. (2.8,2) -- (3,2) {};
	\draw[-] (-1,0) -- (0.8,0) node [near start,above] {\footnotesize{$\mathbf{X}\cap\mathbf{X}'$}};
	\draw[-] (0.8,0) .. controls (1.6,0) and (1.2,0) .. (2,0) -- (3,0) {};
	\draw[-] (-1,-2) -- (0.8,-2) node [near start,above] {\footnotesize{$\mathbf{X}\cap\mathbf{Y'}$}};
	\draw[-] (0.8,-2) .. controls (1.6,-2) and (1.2,-2) .. (2,-2) -- (3,-2) {};
	\draw[-] (-1,-4) -- (0.8,-4) node [near start,above] {\footnotesize{$\mathbf{Y}\cap\mathbf{Y}'$}};
	\draw[-] (0.8,-4) .. controls (1.6,-4) and (1.2,-4) .. (2,-4) -- (3,-4) {};
	\draw[-] (-1,2) -- (0.8,2) node [near start,above] {\footnotesize{$\mathbf{Y}\backslash\mathbf{X}'\cup\mathbf{Y}'$}};
	\draw[-] (0.8,2) .. controls (1,2) and (1.2,1.8) .. (1.2,1) -- (1.2,-5) .. controls (1.2,-5.8) and (1.4,-6) .. (2,-6) -- (3,-6) {};
		
	\draw[-,double] (3,2.5) -- (6,2.5) -- (6,-2.5) -- (3,-2.5) -- (3,2.5) {};
	\node (kappa) at (4.5,0) {\Large{$\kappa_0$}};

	\draw[-] (6,0) -- (9,0) {};
	\draw[-] (6,-2) -- (9,-2) node [midway,above] {\footnotesize{$\mathbf{X'}\cap\mathbf{Y}$}};
	\draw[-] (3,-4) -- (9,-4) {};
	\draw[-] (3,-6) -- (9,-6) {};

	\draw[-,double] (9,-1.5) -- (12,-1.5) -- (12,-6.5) -- (9,-6.5) -- (9,-1.5) {};
	\node (kappa') at (10.5,-4) {\Large{$\kappa_1$}};

	\draw[-] (6,2) .. controls (6.6,2) and (6.8,2.2) .. (6.8,3) .. controls (6.8,3.8) and (7,4) .. (7.4,4) -- (15,4) {};	
		\draw[-] (15,4) -- (16,4) node [near end, above] {\footnotesize{$\mathbf{X'}\backslash\mathbf{X}\cup\mathbf{Y}$}};
	\draw[-] (9,0) -- (15,0) {};
		\draw[-] (15,0) -- (16,0) node [near end, above] {\footnotesize{$\mathbf{X'}\cap\mathbf{X}$}};
	\draw[-] (12,-2) -- (15,-2) {};
		\draw[-] (15,-2) -- (16,-2) node [near end, above] {\footnotesize{$\mathbf{Y'}\cap\mathbf{X}$}};
	\draw[-] (12,-4) -- (15,-4) {};
		\draw[-] (15,-4) -- (16,-4) node [near end, above] {\footnotesize{$\mathbf{Y'}\cap\mathbf{Y}$}};
	\draw[-] (12,-6) .. controls (12.8,-6) and (13,-5.8) .. (13,-5) -- (13,1) .. controls (13,1.8) and (13.2,2) .. (13.6,2) -- (15,2) {};
		\draw[-] (15,2) -- (16,2) node [near end, above] {\footnotesize{$\mathbf{Y}'\backslash\mathbf{X}\cup\mathbf{Y}$}};

	\draw[-] (16,-8) -- (-1,-8) {};
	\draw[-] (16,-4) .. controls (18,-4) and (18,-8) .. (16,-8) {};
	\draw[-] (-1,-4) .. controls (-3,-4) and (-3,-8) .. (-1,-8) {};
	\draw[-] (16,-10) -- (-1,-10) {};
	\draw[-] (16,-2) .. controls (19,-2) and (19,-10) .. (16,-10) {};
	\draw[-] (-1,-2) .. controls (-4,-2) and (-4,-10) .. (-1,-10) {};
	\draw[-] (16,-12) -- (-1,-12) {};
	\draw[-] (16,0) .. controls (20,0) and (20,-12) .. (16,-12) {};
	\draw[-] (-1,0) .. controls (-5,0) and (-5,-12) .. (-1,-12) {};
\end{tikzpicture}
\end{center}
\vspace{1cm}
\begin{center}
\begin{tikzpicture}[x=0.55cm,y=0.38cm]
	\draw[-] (-1,2) -- (0.8,2) node [near start,above] {\footnotesize{$\mathbf{Y}\backslash\mathbf{Y}'\cup\mathbf{X'}$}};
	\draw[-] (0.8,2) -- (3,2) {};
	\draw[-] (-1,0) -- (0.8,0) node [near start,above] {\footnotesize{$\mathbf{Y}\cap\mathbf{Y}'$}};
	\draw[-] (0.8,0) .. controls (1.6,0) and (1.2,0) .. (2,0) -- (3,0) {};
	\draw[-] (-1,-2) -- (0.8,-2) node [near start,above] {\footnotesize{$\mathbf{Y}\cap\mathbf{X'}$}};
	\draw[-] (0.8,-2) .. controls (1.6,-2) and (1.2,-2) .. (2,-2) -- (3,-2) {};
	\draw[-] (-1,-4) -- (0.8,-4) node [near start,above] {\footnotesize{$\mathbf{X}\cap\mathbf{X}'$}};
	\draw[-] (0.8,-4) .. controls (1.6,-4) and (1.2,-4) .. (2,-4) -- (3,-4) {};
	\draw[-] (-1,4) -- (0.8,4) node [near start,above] {\footnotesize{$\mathbf{X}\backslash\mathbf{Y}'\cup\mathbf{X}'$}};
	\draw[-] (0.8,4) .. controls (1.6,4) and (1.8,3.8) .. (1.8,3) -- (1.8,-5) .. controls (1.8,-5.8) and (2,-6) .. (2.4,-6) -- (3,-6) {};
		
	\draw[-,double] (3,2.5) -- (6,2.5) -- (6,-2.5) -- (3,-2.5) -- (3,2.5) {};
	\node (kappa) at (4.5,0) {\Large{$\kappa_1$}};

	\draw[-] (6,2) -- (9,2) {};	
	\draw[-] (6,0) -- (9,0) {};
	\draw[-] (6,-2) -- (9,-2) node [midway,above] {\footnotesize{$\mathbf{Y'}\cap\mathbf{X}$}};
	\draw[-] (3,-4) -- (9,-4) {};
	\draw[-] (3,-6) -- (9,-6) {};

	\draw[-,double] (9,-1.5) -- (12,-1.5) -- (12,-6.5) -- (9,-6.5) -- (9,-1.5) {};
	\node (kappa') at (10.5,-4) {\Large{$\kappa_0$}};

	\draw[-] (9,2) -- (15,2) {};	
		\draw[-] (15,2) -- (16,2) node [near end, above] {\footnotesize{$\mathbf{Y'}\backslash\mathbf{Y}\cup\mathbf{X}$}};
	\draw[-] (9,0) -- (15,0) {};
		\draw[-] (15,0) -- (16,0) node [near end, above] {\footnotesize{$\mathbf{Y'}\cap\mathbf{Y}$}};
	\draw[-] (12,-2) -- (15,-2) {};
		\draw[-] (15,-2) -- (16,-2) node [near end, above] {\footnotesize{$\mathbf{X'}\cap\mathbf{Y}$}};
	\draw[-] (12,-4) -- (15,-4) {};
		\draw[-] (15,-4) -- (16,-4) node [near end, above] {\footnotesize{$\mathbf{X'}\cap\mathbf{X}$}};
	\draw[-] (12,-6) -- (12.6,-6) .. controls (13,-6) and (13.2,-5.8) .. (13.2,-5) -- (13.2,3) .. controls (13.2,3.8) and (13.4,4) .. (14,4)-- (15,4) {};
		\draw[-] (15,4) -- (16,4) node [near end, above] {\footnotesize{$\mathbf{X}'\backslash\mathbf{Y}\cup\mathbf{X}$}};

	\draw[-] (16,-8) -- (-1,-8) {};
	\draw[-] (16,-4) .. controls (18,-4) and (18,-8) .. (16,-8) {};
	\draw[-] (-1,-4) .. controls (-3,-4) and (-3,-8) .. (-1,-8) {};
	\draw[-] (16,-10) -- (-1,-10) {};
	\draw[-] (16,-2) .. controls (19,-2) and (19,-10) .. (16,-10) {};
	\draw[-] (-1,-2) .. controls (-4,-2) and (-4,-10) .. (-1,-10) {};
	\draw[-] (16,-12) -- (-1,-12) {};
	\draw[-] (16,0) .. controls (20,0) and (20,-12) .. (16,-12) {};
	\draw[-] (-1,0) .. controls (-5,0) and (-5,-12) .. (-1,-12) {};
\end{tikzpicture}
\end{center}
\caption{Figure for proof of \autoref{commutexec}\label{fig:commutexec}}
\end{figure*}
\end{proof}

This establishes the existence of a well-defined associative execution, the first ingredient for constructing linear realisability models. Following what was exposed in the first sections, we now define a zeta function associated with pairs of general sub-Markov processes, and show it satisfies the required cocycle property w.r.t. execution.

\begin{defi}
Given two kernels $\kappa, \kappa'$, we define their \emph{zeta-measurement} $\zetameas{\kappa}{\kappa'}$ as the function $\zeta_{\kappa\bullet\kappa'}(z)$.
\end{defi}

\begin{prop}[Cocycle]\label{Markovcocycle}
Given three sub-Markov kernels $\kappa$ on $\mathbf{X}\times \mathbf{X'}$, $\kappa'$ on $\mathbf{Y}\times \mathbf{Y'}$, and $\kappa''$ on $\mathbf{Z}\times \mathbf{Z'}$ in general position:
\[ \zetameas{\kappa}{\kappa'}(z)\zetameas{\kappa\plug \kappa'}{\kappa''}(z) = \zetameas{\kappa'\plug \kappa''}{\kappa}(z)\zetameas{\kappa'}{\kappa''}(z) \]
\end{prop}

\begin{proof}
The proof consists of heavy computations, but without any technical difficulties. The main ingredient is again the geometric adjunction (Equation \ref{geometrictrefoil}). The pictures shown in the proof of \autoref{assocexec2} can be used here to have better insights on the situation. The zeta function quantifies the finite orbits, i.e. the proportion of points that can be reached from themselves by alternating iterations of the involved kernels (weighted by the probabilities of such dynamics occurring). 
The main ingredient of the proof is then that a closed path alternating between $F$, $G$, and $H$ is either a closed path alternating between $F$ and $G$, or a closed path alternating between $H$ and alternating paths between $F$ and $G$. Since the roles of $F$, $G$ and $H$ are symmetric in this statement, we obtain three different splittings of the initial set of closed paths. Now, since zeta functions measure sets of closed paths, these three equal but different expressions yield three different products of two zeta functions. The statement above simply corresponds to stating the equality of two of those. 
\end{proof}

To construct the model of linear logic, we will now follow the usual process. We need to consider not only kernels, but pairs of a kernel and a function. This is used to capture the information about closed paths appearing during the execution, as in the graph case \cite{seiller-goim}.

\subsection{A first model of Linear Logic}\label{sec:firstmodel}

To obtain a model of full linear logic, one has to consider sub-Markov kernels with a set of states. Following a previous construction of a model of second-order linear logic \cite{seiller-goif}, we will represent the set of states by the segment $[0,1]$. 

\begin{defi}
A \emph{proof-object} of support $\measure{X}$ is a pair $\de{f}=(f,F)$ of a function $\complexN\rightarrow\complexN$ and a sub-Markov kernel $F$ on $(\measure{X}\times[0,1])\times(\measure{X}\times[0,1])$.
\end{defi}

We define the operations $(\_)^{\dagger}$ and $(\_)^{\ddagger}$ that will be used throughout the constructions. These operations are meant to ensure that the sets of states of two proof-objects do not interact. Indeed, those should be understood as sets of control states, such as the states of automata. The set of states of a composition is defined as the product of the sets of states of the two objects composed. Given a sub-Markov kernel $F: (\measure{X}\times[0,1])\times(\measure{X}\times[0,1])\rightarrow [0,1]$, we define $(\kappa)^{\dagger}$ and $(\kappa)^{\ddagger}$ as the following sub-Markov kernels $(\measure{X}\times[0,1]\times [0,1])\times(\measure{X}\times[0,1]\times [0,1]) \rightarrow [0,1]$:
\[
\begin{array}{rrcl}
(\kappa)^{\dagger}: 
				& ((x,e,f),(\dot{x},\dot{e},\dot{f})) & \mapsto & \kappa((x,e),(\dot{x},\dot{e}))\identity(f,\dot{f})\\
(\kappa)^{\ddagger}: 
				& ((x,e,f),(\dot{x},\dot{e},\dot{f})) & \mapsto & \kappa((x,f),(\dot{x},\dot{f}))\identity(e,\dot{e}).\\
\end{array}
\]

\begin{defi}\label{zetameasure}
Given two proof objects $\de{f}=(f,\kappa_{\mathrm{F}})$ and $\de{g}=(g,\kappa_{\mathrm{G}})$ we define the \emph{zeta-measurement} as the function: $\zetameas{f}{g}: z\mapsto f(z).g(z).\zeta_{\kappa^{\dagger}_{\mathrm{F}}\bullet\kappa^{\ddagger}_{\mathrm{G}}}(z).$
\end{defi}

\begin{defi}
The execution of two proof objects $\de{f}=(f,\kappa_{\mathrm{F}})$ and $\de{g}=(g,\kappa_{\mathrm{G}})$ of respective supports $\measure{X}$ and $\measure{Y}$ , is defined as the proof-object $\de{f}\plug\de{g}=(f.g.\zetameas{\kappa_{\mathrm{F}}}{\kappa_{\mathrm{G}}},\kappa^{\dagger}_{\mathrm{F}}\plug \kappa^{\ddagger}_{\mathrm{G}})$.
Note that this is a proof-object up to isomorphism between $[0,1]$ and $[0,1]^2$.
\end{defi}

Based on \autoref{assocexec2} and \autoref{commutexec} and the associativity and commutativity of the pointwise product of functions, this notion of execution is associative and commutative.

We now define the orthogonality relation. This follows the construction on graphs in Section \ref{section:graphs}.

\begin{defi}\label{antipode}
An antipode $P$ is a family of functions $\complexN\rightarrow\complexN$. Given two proof objects $\de{f}=(f,\kappa_{\mathrm{F}})$ and $\de{g}=(g,\kappa_{\mathrm{G}})$ of support $\mathbf{X}$, they are orthogonal w.r.t. the antipode $P$ -- denoted $\de{f}\poll[P] \de{g}$ -- if and only if $\zetameas{f}{g}\in P$.
\end{defi}

We now suppose that an antipode has been fixed until the end of this section. We will therefore omit the subscript. We now explain how to construct a model of second order linear logic. We will omit the description of the construction of additive connectives: it follows from earlier work \cite{seiller-goiadd,seiller-goie} in a straightforward manner. 

\begin{defi}
A \emph{type} of support $\mathbf{V}$ is a set $\cond{A}$ of proof-objects of support $\mathbf{V}$ such that there exists a set $B$ of proof-objects with $\cond{A}=B^{\pol}$. Equivalently, a type is a set $\cond{A}$ such that $\cond{A}=\cond{A}^{\pol\pol}$.
\end{defi}

\begin{defi}
For $\cond{A}$, $\cond{B}$ types of disjoint supports, we define:
\begin{align*}
\cond{A\otimes B} &= \{ \de{a}\plug\de{b} \mid \de{a}\in\cond{A}, \de{b}\in\cond{B}\}^{\pol\pol}\\
\cond{A\multimap B} &= \{ \de{f} \mid \forall \de{a}\in\cond{A}, \de{f\plug a}\in\cond{B}\} 
\end{align*}
\end{defi}

A direct consequence of the cocycle property and these definitions is the following property.
For any two types $\cond{A,B}$ with disjoint support:
\[ \cond{(A\otimes B^{\pol})}^{\pol} = \cond{A\multimap B}. \]

Now, to define exponentials, one has to restrict to specific spaces. Indeed, not all sub-Markov kernels can be exponentiated. This is easy to understand: if a proof-object uses several copies of its argument, it uses it through its set of states. To understand how states allow for this, consider two automata that are composed. If the first automata has two states, it can ask the first automata to perform a computation, change state, and then ask again, triggering two computations of the second machine. This works perfectly \emph{provided the second machine ends its computation on its initial state}, otherwise it would not run correctly the second time as there is no way to reinitiate it. This issue is dealt with in the models by exponentiation, as only exponentiated processes can be used multiple times. To ensure the latter end their computation in the same state as they started, exponentiation replaces the program $\de{a}$ by a single-state program $\de{!a}$, encoding the states of $\de{a}$ in the configuration space to avoid information loss. This encoding requires the underlying space $\mathbf{X}$ to be large enough, i.e. contain the space $[0,1]^\naturalN$. Exponentiation, represented in this way, is therefore defined as long as the underlying space $\mathbf{X}$ contains $[0,1]^\naturalN$.

We thus restrict in this section to spaces of the form $\mathbf{X}=\mathbf{Y}\times[0,1]^\naturalN$, but we will show in Section \ref{section:markovmodel} how to bypass this restriction.

\begin{defi}\label{def:balanced}
A proof-object $(f,\kappa_{\mathrm{F}})$ is \emph{balanced} if $f=1$, the constant function equal to $1$. If $E$ is a set of proof-objects, we write $\balanced{E}$ the subset of balanced proof-objects in $E$.
\end{defi}

Following an earlier model \cite{seiller-goif}, we will define the exponential through the following maps for all space $\mathbf{X}$ as above:
\[
\begin{array}{rrcl}
B_{\measure{X}}:&\mathbf{Y}\times[0,1]^\naturalN\times[0,1]&\rightarrow&\mathbf{Y}\times[0,1]^\naturalN\\
					&(a,s,d)&\mapsto&(a,d: s)
\end{array}
\]
where $:$ denotes here the concatenation.
This map is used to define $\oc \kappa$ from a sub-Markov kernel $\kappa:(\measure{X}\times[0,1])\times(\measure{X}\times[0,1])\rightarrow [0,1]$ (we recall that the copies of $[0,1]$ here represent the set of states of the proof-object). We first define\footnote{Here $B$ is a bijective map, and not a kernel, but we implicitly use the kernel composition by considering the kernel form of $B$ and $B^{-1}$.} $B_{\measure{X}}^{-1}\circ\kappa\circ B_{\measure{X}}$, which is a sub-Markov kernel $\measure{X}\times\measure{X}\rightarrow [0,1]$, and then $\oc\kappa: (\measure{X}\times[0,1])\times(\measure{X}\times[0,1])\rightarrow [0,1]$ can be defined as:
$
\oc \kappa: (x,e,\dot{x},\dot{e}) \mapsto B_{\measure{X}}^{-1}\circ\kappa\circ B_{\measure{X}}(x,\dot{x})\identity(e,\dot{e}).
$
Note that the information of the states of $\kappa$ is encoded in $\oc\kappa$ within the space $\mathbf{X}$ and the latter acts on the set of states as the identity, i.e. as if it has a single state.

\begin{defi}[Perennisation]
Let $\de{f}=(1,\kappa_{\mathrm{F}})$ be a balanced proof-object. We define its \emph{perennisation} $\de{\oc f}=(1,\oc \kappa_{\mathrm{F}})$.
\end{defi}

\begin{defi}[Exponential]
Let $\cond{A}$ be a type. We define the perennial type $\cond{\oc A}$ as the bi-orthogonal closure $\cond{\oc A}=\cond{(\sharp A)^{\pol\pol}}$ where $\cond{\sharp A}$ is the set
$\cond{\sharp A}=\{\de{\oc a}~|~\de{a}\in\balanced{\cond{A}}\}.$
\end{defi}

This defines a model of second-order linear logic (\secLL) using the constructions from the author's work on graphings \cite{seiller-goif}.
\begin{thm}
Restricting to spaces $\mathbf{X}=\mathbf{Y}\times[0,1]^\naturalN$, proof-objects and types define a sound model of \secLL.
\end{thm}

\begin{proof}
In the \IG model for full linear logic \cite{seiller-goif}, linear logic proofs are interpreted by \emph{deterministic graphings}. As such, they are in fact interpreted by dynamical systems by \autoref{detdynsys}, which in turn define sub-Markov kernels.
\end{proof}

The model just sketched restricts the type of spaces considered. We will therefore devote the next section to bypass this issue, explaining how to model \secLL in the unrestricted setting of sub-Markov kernels by defining a new interpretation of exponential connectives.

\section{Sub-Markov processes and linear logic \label{section:markovmodel}}

\begin{notations}
When writing down explicit formulas for the value of a sub-kernel $\kappa$ on $(\mathbf{X}\times[0,1])\times(\mathbf{Y}\times[0,1])$, we will notationally separate the set of states and the spaces $\mathbf{X}$ and $\mathbf{Y}$. That is, we will write $\kappa: \mathbf{X}\cdot[0,1] \times \mathbf{Y}\cdot [0,1]$ and write explicit definitions as $\kappa(x;\dot{y})\cdot(e;\dot{f})$ to denote $\kappa(x,e,\dot{y},\dot{f})$.
\end{notations}

To avoid restricting to spaces of the form $\mathbf{X}=\mathbf{Y}\times[0,1]^\naturalN$, we will consider that $\kappa$ and $\oc\kappa$ need not act on the same space: while $\kappa$ is defined on $\mathbf{X}\times\mathbf{Y}$, $\oc\kappa$ will be defined on $(\mathbf{X}\times[0,1])\times(\mathbf{Y}\times[0,1])$. This implies that we need to generalise the framework to define proof-objects with an underlying sub-Markov kernel on $\mathbf{X}\times\mathbf{Y}$ and not necessarily on $\mathbf{X}\times\mathbf{X}$.

\begin{notations}
In the following, when considering proof-objects $(f,\kappa_\mathrm{F})$, we will say $\kappa_\mathrm{F}$ is a sub-Markov kernel \emph{from $\mathbf{X}$ to $\mathbf{Y}$} to express that $\kappa_\mathrm{F}$ has type $\mathbf{X}\cdot[0,1]\times\mathbf{Y}\cdot[0,1]\rightarrow[0,1]$. 
\end{notations}

We now detail this construction, which requires redefining parts of the interpretations of linear logic proofs from \cite{seiller-goif}. 

\subsection{Multiplicatives}
The definition of orthogonality, types, and multiplicative connectives follow the constructions exposed in previous sections.

\begin{defi}
A \emph{general proof-object} of support $\mathbf{X}\rightarrow\mathbf{Y}$ is a pair $\de{f}=(f,\kappa_\mathrm{F})$ of a complex function $f$ and a sub-Markov kernel $\kappa_\mathrm{F}$ from $\mathbf{X}$ to $\mathbf{Y}$.
\end{defi}

The zeta-measurement and the notion of antipode are defined as above (\autoref{zetameasure} and \autoref{antipode}).

\begin{defi}
Two general proof-objects $\de{f},\de{g}$ of respective supports $\mathbf{X}\rightarrow\mathbf{Y}$ and $\mathbf{Y}\rightarrow\mathbf{X}$ are orthogonal w.r.t. an antipode $P$, which is denoted by $\de{f}\poll[P]\de{g}$, when $\zetameas{f}{g}\in P$.
\end{defi}

From now on, we fix an antipode and omit subscripts.

\begin{defi}
A \emph{type} of support $\mathbf{V}$ is a set $\cond{A}$ of general proof-objects of support $\mathbf{V}$ such that $\cond{A}=\cond{A}^{\poll{}\poll{}}$.
\end{defi}

\begin{defi}
Given two general proof-objects $\de{f},\de{g}$ of respective supports $\mathbf{X}\rightarrow\mathbf{Y}$ and $\mathbf{X}'\rightarrow\mathbf{Y}'$, their execution is the proof-object of support 
$ (\mathbf{X}\cup\mathbf{X}')\backslash(\mathbf{Y}\cup\mathbf{Y}')\rightarrow(\mathbf{Y}\cup\mathbf{Y}')\backslash(\mathbf{X}\cup\mathbf{X}')$ defined as $\de{f}\plug\de{g}=(\zetameas{f}{g},\kappa_\mathrm{F}\plug\kappa_\mathrm{G})$.
\end{defi}

\begin{rem}
Notice that the execution is not commutative here. Commutativity can be shown as long as one requires that $\mathbf{X}\cap\mathbf{X}'$ and $\mathbf{Y}\cap\mathbf{Y}'$ are negligible.
\end{rem}

\begin{defi}
Let $\de{f},\de{g}$ be two general proof-objects of respective supports $\mathbf{X}\rightarrow\mathbf{Y}$ and $\mathbf{X}'\rightarrow\mathbf{Y}'$, where $\mathbf{X},\mathbf{X}',\mathbf{Y},\mathbf{Y}'$ are pairwise disjoint. We write $\de{f}\otimes\de{g}$ for the execution of $\de{f}$ and $\de{g}$.
\end{defi}

Note that $\de{f}\otimes\de{g}=\de{g}\otimes\de{f}$ by the above remark.

\begin{notations}
We say that two types $\cond{A},\cond{B}$ of support $\mathbf{X}\rightarrow\mathbf{Y}$ and $\mathbf{X}'\rightarrow\mathbf{Y}'$ are of \emph{disjoint support} when $\mathbf{X},\mathbf{X}',\mathbf{Y},\mathbf{Y}'$ are pairwise disjoint.
\end{notations}

\begin{defi}
Let $\cond{A},\cond{B}$ be types of disjoint supports $\mathbf{X}\rightarrow\mathbf{Y}$ and $\mathbf{X}'\rightarrow\mathbf{Y}'$. We define
\begin{align*}
\cond{A\otimes B}&=\{\de{a}\otimes\de{b}\mid \de{a}\in\cond{A}, \de{b}\in\cond{B}\}^{\poll{}\poll{}}\\
\cond{A\multimap B}&=\{\de{f}\mid \forall \de{a}\in\cond{A}, \de{f}\plug\de{a}\in\cond{B}\}
\end{align*}
of respective supports $\mathbf{X}\times\mathbf{X'}\rightarrow\mathbf{Y}\times \mathbf{Y}$ and $\mathbf{Y}\times\mathbf{X'}\rightarrow\mathbf{X}\times \mathbf{Y}$.
\end{defi}

The following theorem then establishes that multiplicative connectives are adequately interpreted. We omit the proof which is standard \cite{seiller-goim,seiller-goiadd,seiller-goig}.
\begin{thm}
Let $\cond{A}$ and $\cond{B}$ be types of disjoint supports $\mathbf{X}\rightarrow\mathbf{Y}$ and $\mathbf{X}'\rightarrow\mathbf{Y}'$.  We have:
\[ \cond{A\multimap B} = (\cond{A\otimes B^{\poll{}}})^{\poll{}{}}\]
\end{thm}

\subsection{Additives and Quantifiers}

To represent additives, one uses the notion of state: the additive conjunction $\with$ superposes two proof-objects of the same support whose sets of states $S$ and $S'$ by creating a proof-object whose set of states is $S+S'$. 
\begin{notations}
We write $\de{0}_{\mathbf{X}\rightarrow\mathbf{Y}}$ for the proof-object $(1,\mathbf{0}_{\mathbf{X}\rightarrow\mathbf{Y}})$ where $1$ is a the constant function equal to $1$ and $\mathbf{0}_{\mathbf{X}\rightarrow\mathbf{Y}}$ is the zero sub-Markov kernel from $\mathbf{X}$ to $\mathbf{Y}$, i.e. $\mathbf{0}(x,\_)$ is the constant $0$ subprobability distribution.
\end{notations}

\begin{notations}
Let $A,B$ be kernels respectively from $\mathbf{X}$ to $\mathbf{Y}$ and from $\mathbf{X}$ to $\mathbf{Y}$. We write $A\with B$ for the kernel $\kappa$ from $\mathbf{X}$ to $\mathbf{Y}$ defined as $\kappa(x,y)\cdot(e,\dot{f})=A(x,y)\cdot(2e,\dot{f})$ if $0\leqslant e \leqslant 1/2$ and $\kappa(x,y)\cdot(e,\dot{f})=B(x,y)\cdot(2e-1,\dot{f})$ otherwise.
\end{notations}

\begin{defi}
If $\de{a}=(a,A)$ and $\de{b}=(b,B)$ are proof-objects of support $\mathbf{X}\rightarrow\mathbf{Y}$, we define $\de{a\with b}=(a+b,A\with B)$ of support $\mathbf{X}\rightarrow\mathbf{Y}$.
\end{defi}

Additives are defined on a subset of types that never allow for weakening called \emph{behaviour} in earlier work \cite{seiller-goie}; we here use the terminology \emph{purely linear types}.

\begin{defi}
A type $\cond{A}$ has the expansion property when $\forall \de{a}\in\cond{A}, \de{a}\with \de{0}_{\mathbf{X}\rightarrow\mathbf{Y}} \in\cond{A}$. A type $\cond{A}$ is \emph{purely linear} if both $\cond{A}$ and $\cond{A}^{\poll{}{}}$ have the expansion property.
\end{defi}

\begin{defi}
Let $\cond{A},\cond{B}$ be purely linear types of disjoint supports $\mathbf{X}\rightarrow\mathbf{Y}$ and $\mathbf{X}'\rightarrow\mathbf{Y}'$. We define the following purely linear types of support $\mathbf{X}\times\mathbf{X'}\rightarrow \mathbf{Y}\times \mathbf{Y'}$:
\begin{align*}
\cond{A\oplus B} &= (\{\de{a}\otimes\de{0}_{\mathbf{X'}\rightarrow\mathbf{Y'}}\mid \de{a}\in\cond{A}\}\cup\{\de{0}_{\mathbf{X}\rightarrow\mathbf{Y}}\otimes\de{b}\mid \de{b}\in\cond{B}\})^{\poll{}{}\poll{}{}},\\
\cond{A\with B} &= \{(\de{a}\otimes\de{0}_{\mathbf{X'}\rightarrow\mathbf{Y'}})\with(\de{0}_{\mathbf{X}\rightarrow\mathbf{Y}}\otimes\de{b}) \mid \de{a}\in\cond{A}, \de{b}\in\cond{B}\}^{\poll{}{}\poll{}{}}.
\end{align*}
\end{defi}

\begin{defi}
We define (support-wise) second-order quantification as the following operations on types (not necessarily purely linear):
\begin{align*}
\mathbb{\forall}_{\mathbf{X}\rightarrow\mathbf{Y}} \cond{X~ F(X)} &= \bigcap_{\cond{A}\text{ of support }\mathbf{X}\rightarrow\mathbf{Y}} \cond{F(A)}\\
\mathbb{\exists}_{\mathbf{X}\rightarrow\mathbf{Y}} \cond{X~ F(X)} &= \left(\bigcup_{\cond{A}\text{ of support }\mathbf{X}\rightarrow\mathbf{Y}} \cond{F(A)}\right)^{\poll{}{}\poll{}{}}
\end{align*}
\end{defi}

These definitions of additives and quantifiers, together with the interpretation of multiplicatives explained in the previous section, allow us to interpret second-order multiplicative additive linear logic (\malltwo); proofs follow closely the author's previous work \cite{seiller-goig}.

\subsection{Exponentials}
We redefine exponentials for balanced proof-objects.

\begin{defi}
Let $\de{f}=(1,\kappa)$ be a balanced general proof-object of support $\mathbf{X}\rightarrow \mathbf{Y}$. We define $\oc \de{f}$ as the general proof-object $(1,\oc\kappa)$ where $\oc\kappa$ is the sub-Markov process from $\mathbf{X}\times[0,1]$ to $\mathbf{Y}\times[0,1]$ defined as:
\[
\oc\kappa(x,e;\dot{x},\dot{e})\cdot(f;\dot{f})= \kappa(x;\dot{x})\cdot(e;\dot{e})\identity(f;\dot{f})
\]
\end{defi}

\begin{defi}[Exponentiation]
Let $\de{f}=(1,\kappa_{\mathrm{F}})$ be a balanced proof-object. We define its \emph{exponential} as $\de{\oc f}=(1,\oc \kappa_{\mathrm{F}})$.
\end{defi}

We will now show that the exponential principles of linear logic can be interpreted faithfully. 
\begin{notations}
Given a measurable map $f: \mathbf{X} \rightarrow \mathbf{Y}$, it induces a kernel $\kappa_f$ on $\mathbf{X}\times\mathbf{Y}$ defined as $\kappa_f(x,\dot{y})=\identity(f(x),\dot{y})$. It also induces a kernel $\kappa_f^\ast$ on $\mathbf{Y}\times\mathbf{X}$ defined as $\kappa_f^\ast(y,\dot{x})=\identity(f(x),\dot{y})$. Note that if $f$ is bijective, $\kappa_f^\ast=\kappa_{f^{-1}}$.

We will also use the sum symbol $+$ to denote the parallel composition of kernels, i.e. given kernels $\kappa$ on $\mathbf{X}\times\mathbf{Y}$ and $\kappa'$ on $\mathbf{Z}\times\mathbf{W}$, the kernel $\kappa+\kappa'$ on $(\mathbf{X}+\mathbf{Z})\times(\mathbf{Y}+\mathbf{W})$ is defined as 
\[
(u,v)\mapsto \left\{\begin{array}{ll}
	\kappa(u,\dot{v}) & \text{if $u\in\mathbf{X}$, $\dot{v}\in\mathbf{Y}$}\\
	\kappa'(u,\dot{v}) & \text{if $u\in\mathbf{Z}$, $\dot{v}\in\mathbf{W}$}\\
	0 & \text{otherwise}\\
\end{array}\right.
\]

Lastly, if $\kappa$ is a kernel on $\mathbf{X}\times\mathbf{Y}$, we write $\bar{\kappa}$ the kernel $\kappa$ extended with a dialect on which it acts as the identity, i.e. $\bar{\kappa}$ is the kernel on $(\mathbf{X}\times[0,1])\times(\mathbf{Y}\times[0,1])$ (i.e. the kernel \emph{from $\mathbf{X}$ to $\mathbf{Y}$}) defined as $\bar{\kappa}(x;\dot{y})\cdot(e;\dot{f})=\kappa(x,\dot{y})\identity(e,\dot{f})$.
\end{notations} 

The following lemma, established by the author \cite[Proposition 37]{seiller-goie}, will be particularly useful in the following proofs. It states that to prove a proof-object $\de{f}$ belongs to $\cond{A}\multimap\cond{B}$, it is enough to prove $\de{f}\plug\de{a}\in\cond{B}$ when $\de{a}$ ranges over a \emph{generating} set for $\cond{A}$.

\begin{lem}\label{claimethicmaps}
Let $\cond{A,B}$ be types and $E$ a generating set for $\cond{A}$, i.e. $\cond{A}=E^{\pol\pol}$. If $\de{f}$ is such that $\forall \de{a}\in\cond{A}, \de{f}\plug\de{a}\in\cond{B}$, then $\de{f}$ belongs to the type $\cond{A\multimap B}$.
\end{lem}

\begin{prop}\label{prop:digging}
The digging rule can be interpreted.
\end{prop}

\begin{proof}
Suppose $\cond{A}$ is of support $\mathbf{X}\rightarrow\mathbf{Y}$. This map is easily implemented as a project $\de{dig}_\mathbf{X\rightarrow Y}=(1,\kappa_{\mathrm{dig}_\mathbf{Y}}+\kappa_{\mathrm{dig}_\mathbf{X}}^{\ast})$ with
\begin{equation*}
\begin{array}{rr}
\kappa_{\mathrm{dig}_\mathbf{X}}^{\ast}:
&(\mathbf{X}\times[0,1]\times[0,1])\cdot[0,1] \times (\mathbf{X}\times[0,1])\cdot[0,1]\rightarrow [0,1]\\
&(x,e,e')\cdot e'',(x',f)\cdot f'\mapsto \identity(x,x')\identity(e,f)\identity(\varphi(e',e''),f')
\end{array}
\end{equation*}
\begin{equation*}
\begin{array}{rr}
\kappa_{\mathrm{dig}_\mathbf{Y}} : 
&(\mathbf{Y}\times[0,1])\cdot[0,1] \times (\mathbf{Y}\times[0,1]\times[0,1])\cdot[0,1]\rightarrow [0,1]\\
&(x,e)\cdot e',(x',f,f')\cdot f''\mapsto \identity(x,x')\identity(e,f)\identity(e',\varphi(f',f''))
\end{array}
\end{equation*}
where $\varphi$ is a fixed bijection between $[0,1]$ and $[0,1]^2$.

To show that this implements digging is a simple exercise. Taking $\de{a}$ a balanced project, we consider $\oc\de{a}$, and show that $\oc\de{a}\plug\de{dig}_\mathbf{X\rightarrow Y}=\oc\oc\de{a}$. In fact, it is easy to convince oneself that the kernel of $\oc\de{a}\plug\de{dig}_\mathbf{X\rightarrow Y}$ is computed as the set of \enquote{length 3 paths}. 
The following computation proves that $\oc\de{a}\plug\de{dig}_\mathbf{X\rightarrow Y}=\oc\oc\de{a}$ (up to the isomorphism $\varphi$ between $[0,1]^2$ and $[0,1]$ on the stateset):
\begin{eqnarray*}
\lefteqn{\tr(\oc\kappa_{\mathrm{A}}^{\dagger}\bullet (\kappa_{\mathrm{dig}_\mathbf{Y}}+\kappa^{\ast}_{\mathrm{dig}_\mathbf{X}})^{\ddagger})((x,e,f);(\dot{x},\dot{e},\dot{f}))\cdot(g,h;\dot{g},\dot{h}) }\\
	&=& \int_{(y,u,d,d'),(z,v,c,c')} 
	\left[\begin{array}{l}
			\kappa^{\ast}_{\mathrm{dig}_\mathbf{X}}((x,e,f),(\dot{y},\dot{u}))\cdot(g,\dot{d})\identity(h,\dot{d}')\\
			\hspace{2em}\times\oc\kappa_{\mathrm{A}}((y,u);(\dot{z},\dot{v}))\cdot(d',\dot{c}')\identity(d,\dot{c})\\
			\hspace{4em}\times\kappa_{\mathrm{dig}_\mathbf{Y}}((z,v),(\dot{x},\dot{e},\dot{f}))\cdot(c,\dot{g})\identity(c',\dot{h})
		\end{array}\right]\\
	&=& \int_{(y,u,d,d')} 
	\left[\begin{array}{l}
			\identity(x,\dot{y})\identity(e,\dot{u})\identity(\varphi(f,g),\dot{d})\identity(h,\dot{d}')\\
			\hspace{2em}\times\int_{(z,v,c,c')} \left[\begin{array}{l}
			\oc\kappa_{\mathrm{A}}((y,u);(\dot{z},\dot{v}))\cdot(d',\dot{c}')\identity(d,\dot{c})\\
			\hspace{2em}\times\kappa_{\mathrm{dig}}((z,v),(\dot{x},\dot{e},\dot{f}))\cdot(c,\dot{g})\identity(c',\dot{h})
		\end{array}\right]
		\end{array}\right]\\
	&=& \int_{(z,v,c,c')} 
			\oc\kappa_{\mathrm{A}}((x,e);(\dot{z},\dot{v}))\cdot(h,\dot{c}')\identity(\varphi(f,g),\dot{c})
			\times\kappa_{\mathrm{dig}}((z,v),(\dot{x},\dot{e},\dot{f}))\cdot(c,\dot{g})\identity(c',\dot{h})
\\
	&=& \int_{(z,v,c,c')} 	
			\oc\kappa_{\mathrm{A}}((x,e);(\dot{z},\dot{v}))\cdot(h,\dot{e})\identity(\varphi(f,g),\dot{e}')
			\times\identity(z,\dot{x})\identity(v,\dot{e})\identity(c,\varphi(\dot{f},\dot{g}))\identity(c',\dot{h})
\\	
	&=& \oc\kappa_{\mathrm{A}}((x,e);(\dot{x},\dot{e}))\cdot(h,\dot{h})\identity(\varphi(f,g),\varphi(\dot{f},\dot{g}))\\
	&=& \oc\kappa_{\mathrm{A}}((x,e);(\dot{x},\dot{e}))\cdot(h,\dot{h})\identity(f,\dot{f})\identity(g,\dot{g})\\
	&=& \kappa_{\mathrm{A}}(x;\dot{x})\cdot(e;\dot{e})\identity(h,\dot{h})\identity(f,\dot{f})\identity(g,\dot{g})\\
	&=& \kappa_{\mathrm{A}}(x;\dot{x})\cdot(e;\dot{e})\identity(f,\dot{f})\identity(\varphi(h,g),\varphi(\dot{h},\dot{g}))\\
	&=& \oc\kappa_{\mathrm{A}}((x,e);(\dot{x},\dot{e}))\cdot(f,\dot{f})\identity(\varphi(h,g),\varphi(\dot{h},\dot{g}))\\
	&=& \oc\oc\kappa_\mathrm{A}((x,e,f);(\dot{x},\dot{e},\dot{f}))\cdot(\varphi(h,g),\varphi(\dot{h},\dot{g}))
\end{eqnarray*}

This ends the proof: since $\oc\cond{A}$ is generated by the elements of the form $\oc\de{a}$, the fact that $\de{dig}_\mathbf{X\rightarrow Y}$ maps every element of the form $\oc\de{a}$ to an element of $\oc\oc\cond{A}$ suffices to establish that it belongs to $\oc\cond{A}\multimap\oc\oc\cond{A}$, by Lemma \ref{claimethicmaps}.
\end{proof}

\begin{prop}\label{prop:deriliction}
The dereliction rule can be interpreted.
\end{prop}

\begin{proof}
Now, dereliction is a map $\oc A \multimap A$. Suppose $\cond{A}$ is of support $\mathbf{X}\rightarrow\mathbf{Y}$. This map is easily implemented as a project $\de{der}=(1,\kappa^{\ast}_{\mathrm{der}_\mathbf{X}}+\kappa_{\mathrm{der}_\mathbf{Y}})$ with:
\begin{equation*}
\begin{array}{rr}
\kappa^{\ast}_{\mathrm{der}_\mathbf{X}} : 
&(\mathbf{X})\cdot[0,1] \times \mathbf{X}\times[0,1]\cdot[0,1]\rightarrow [0,1]\\
&x\cdot e,(x',f)\cdot f'\mapsto \identity(x,x')\identity(e,\varphi(f,f))
\end{array}
\end{equation*}
\begin{equation*}
\begin{array}{rr}
\kappa_{\mathrm{der}_\mathbf{Y}} :
&(\mathbf{Y}\times[0,1])\cdot[0,1] \times \mathbf{Y}\cdot[0,1]\rightarrow [0,1]\\
&(x,e)\cdot e',x'\cdot f\mapsto \identity(x,x')\identity(\varphi(e,e'),f)
\end{array}
\end{equation*}
where $\varphi$ is a fixed bijection between $[0,1]$ and $[0,1]^2$.

To show that this implements dereliction is a simple computation. Taking $\de{a}$ a balanced project, we consider $\oc\de{a}$, and show that $\oc\de{a}\plug\de{der}_{\mathbf{X}\rightarrow\mathbf{Y}}=\de{a}$ up to some bijection on the stateset. Now, we compute $\oc\de{a}\plug\de{der}$ with $\de{\oc a}=(1,\oc \kappa)$. Again, given the definition of $\de{der}$, this consists in computing paths of length 3.
The following computation shows that $\oc\de{a}\plug\de{der}_{\mathbf{X}\rightarrow\mathbf{Y}}=\de{a}$ up to the isomorphism between $[0,1]^3$ and $[0,1]^2$ defined by $(a,b,c)\mapsto(\varphi(a,c),b)$:
\begin{eqnarray*}
\lefteqn{\tr(\oc\kappa_{\mathrm{A}}^{\dagger}\bullet (\kappa^{\ast}_{\mathrm{der}_\mathbf{X}}+\kappa_{\mathrm{der}_\mathbf{Y}})^{\ddagger})(x;\dot{x})\cdot(e,f;\dot{e};\dot{f}) }\\
&=& \int_{(y,a,b,c)} \int_{(z,u,v,w)} 
  \left[\begin{array}{l}
			\kappa^{\ast}_{\mathrm{der}_\mathbf{X}}(x;(\dot{y},\dot{a}))\cdot(e;\dot{b})\identity(f,\dot{c})\\ 
			\hspace{2em}\times \oc\kappa_{\mathrm{A}}((y,a);(\dot{z},\dot{u}))\cdot(c;\dot{w})\identity(b;\dot{v})\\
			\hspace{4em}\times \kappa_{\mathrm{der}_\mathbf{Y}}((z,u);\dot{x})\cdot(v;\dot{e})\identity(w,\dot{f})
		\end{array}\right]\\
&=& \int_{(y,a,b,c)} \identity(x;\dot{y})\identity(e;\varphi(\dot{a},\dot{b}))\identity(f,\dot{c}) 
\int_{(z,u,v,w)}  \left[\begin{array}{l}
			\oc\kappa_{\mathrm{A}}((y,a);(\dot{z},\dot{u}))\cdot(c;\dot{w})\identity(b;\dot{v})\\
			\hspace{2em}\times \kappa_{\mathrm{der}_\mathbf{Y}}((z,u);\dot{x})\cdot(v;\dot{e})\identity(w,\dot{f})
		\end{array}\right]\\	
&=& \int_{(z,u,v,w)}  \left[
			\oc\kappa_{\mathrm{A}}((x,\varphi^{-1}_0(e));(\dot{z},\dot{u}))\cdot(f;\dot{w})\identity(\varphi^{-1}_1(e);\dot{v})
			\times \kappa_{\mathrm{der}_\mathbf{Y}}((z,u);\dot{x})\cdot(v;\dot{e})\identity(w,\dot{f})
		\right]\\	
&=& \int_{(z,u,v,w)}  \left[
			\oc\kappa_{\mathrm{A}}((x,\varphi^{-1}_0(e));(\dot{z},\dot{u}))\cdot(f;\dot{w})\identity(\varphi^{-1}_1(e);\dot{v})
			\times \identity(z;\dot{x})\identity(\varphi(u,v);\dot{e})\identity(w,\dot{f})
		\right]\\		
&=& 		\oc\kappa_{\mathrm{A}}((x,\varphi^{-1}_0(e));(\dot{x},\varphi^{-1}_0(\dot{u})))\cdot(f;\dot{f})\identity(\varphi^{-1}_1(e);\varphi^{-1}_1(\dot{v}))\\	
&=&		\kappa_{\mathrm{A}}(x;\dot{x})\cdot(\varphi^{-1}_0(e);\varphi^{-1}_0(\dot{e}))\identity(f,\dot{f})\identity(\varphi^{-1}_1(e);\varphi^{-1}_1(\dot{v}))	
\end{eqnarray*}

Again, by Claim \ref{claimethicmaps} this is enough to establish that $\de{der}_{\mathbf{X}\rightarrow\mathbf{Y}}$ belongs to $\oc\cond{A}\multimap\cond{A}$.
\end{proof}

\begin{prop}\label{prop:promotion}
Functorial promotion holds.
\end{prop}

\begin{proof}
The proof is a tad more involved than the preceding ones. Here we will implement the rule in three steps. The principle is easy to understand: given $\oc\de{a}\in\oc\cond{A}$ and $\oc\de{f}\in\oc(\cond{A}\multimap\cond{B})$, we will first compute the executions $\oc\de{a}\plug\de{left}$ and $\oc\de{f}\plug\de{right}$ in order to ensure disjointness of the spaces used to encode the statesets of $\de{a}$ and $\de{f}$ respectively. Once this is done, the execution $(\oc\de{a}\plug\de{left})\plug(\oc\de{f}\plug\de{right})$ morally computes the same as $\oc\de{f\plug a}$ up to some transformation $\de{fit}$ that internalises a stateset isomorphism. 

Let $\de{a}\in\cond{A}$ and $\de{f}\in\cond{A\multimap B}$ be balanced proof-objects, of respective supports $\mathbf{X}\rightarrow \mathbf{Y}$ and $\mathbf{X}'\rightarrow\mathbf{Y}'$. We consider the proof-object $\de{twist}=(1,\kappa_\mathrm{twist})$ with:
\begin{equation*}
\begin{array}{rr}
	\kappa_\mathrm{twist}: 
	&		(\mathbf{X}\cup\mathbf{X}')\times[0,1]\cdot[0,1]  (\mathbf{Y}\cup\mathbf{Y}')\times[0,1]\cdot[0,1] \rightarrow  [0,1]\\
			& (x,e)\cdot f  (\dot{y},\dot{e})\cdot\dot{f}  \mapsto  \identity(x,\dot{y})\identity(e,\varphi^{-1}_0(\dot{f}))\identity(f,\varphi(\dot{e},\varphi^{-1}(\dot{f}))
\end{array}
\end{equation*}
for $x,\dot{y}\in\mathbf{Y}\cap\mathbf{X}$.
We then use the kernels:
\begin{equation*}
\begin{array}{rr}
	\kappa_{\texttt{l}} : 
&(\mathbf{X}\times[0,1])\cdot[0,1] \times (\mathbf{X}\times[0,1]\times[0,1])\cdot[0,1]\rightarrow [0,1]\\
&(x,e)\cdot e',(x',f,f')\cdot f''\mapsto \identity(x,x')\identity(e,f)\identity(e',\varphi(f',f''))
\end{array}
\end{equation*}
\begin{equation*}
\begin{array}{rr}
\kappa_{\texttt{r}} : 
&(\mathbf{X}\times[0,1])\cdot[0,1] \times (\mathbf{X}\times[0,1]\times[0,1])\cdot[0,1]\rightarrow [0,1]\\
&(x,e)\cdot e',(x',f,f')\cdot f''\mapsto \identity(x,x')\identity(e,f')\identity(e',\varphi(f,f''))
\end{array}
\end{equation*}
\begin{equation*}
\begin{array}{rr}
\kappa_{\texttt{c}} : 
&(\mathbf{X}\times[0,1]\times[0,1])\cdot[0,1] \times (\mathbf{X}\times[0,1])\cdot[0,1]\rightarrow [0,1]\\
&(x,e,e')\cdot e'',(x',f)\cdot f'\mapsto \identity(x,x')\identity(\varphi(e,e'),f)\identity(e',f')
\end{array}
\end{equation*}
where $\varphi$ is a fixed bijection between $[0,1]$ and $[0,1]^2$.

We now consider projects $\de{\oc a}=(1,\oc \kappa_\mathrm{A})$ and $\de{\oc f}=(1,\oc \kappa_\mathrm{F})$ of support $\mathbf{X}$ and $\mathbf{X\times Y}$ respectively. A computation similar to those from the proofs of Proposition \ref{prop:digging} and \ref{prop:deriliction} (but more involved, as paths are not limited to length 3 here) show that 
\[ ((\oc \kappa_\mathrm{A}\plug \kappa_{\texttt{l}})\plug \kappa_\mathrm{twist}\plug(\oc \kappa_\mathrm{F}\plug \kappa_{\texttt{r}}))\plug \kappa_{\texttt{c}}\]
is equal to $\oc \kappa_\mathrm{A\plug F}$.
\end{proof}

We now have stated all the key results needed to establish our main theorem, adapting a previous interpretation of second-order linear logic sequent calculus \cite{seiller-goif}. Similarly to what was shown in the graphing case, it is possible to interpret second-order linear logic formulas $A$ as types $\Interpret{A}$, and proofs of a formula $A$ as proof objects belonging to $\Interpret{A}$. These interpretations can be shown to be invariant under cut-elimination (which is represented by the execution of proof objects) up to observational equivalence, yielding a denotational model.

\begin{thm}
General proof-objects and types define a sound model of \secLL.
\end{thm}

\section{Perspectives}

We established that sub-Markov processes provide a model of second-order linear logic. Probabilistic languages with sampling instructions should be interpretable in this model. For instance, while axiom rules are interpreted by the identity kernel -- i.e. the Dirac delta function --, generalised rules introducing non-trivial Markov kernels can very well be considered. We expect strong connections with game semantics models dealing with such languages \cite{CastellanPaquet}, although our approach differs from the start by its intention. In particular, the realisability approach provides a very rich notion of types arising from the behaviour of processes. This can incorporate dependent types \cite{goi5}, and could be used to consider new type constructions adapted to probabilistic computation \cite{seiller-axioms}. 

It is also worth noting that the formal relation with zeta functions could turn out to be of great interest with respect to the recasting of complexity theory by means of Interaction Graphs models \cite{seiller-towards,seiller-goinda}. Indeed, it is hoped that invariants from dynamical systems (and the group/monoid action used to restrict graphings) to be related to the expressivity of the models, and Seiller and Pellissier established using the framework of graphings that strong algebraic lower bounds can be obtained using \emph{topological entropy} \cite{seiller-PRAMsLB}. The current work thus provides an additional element with respect to these ideas, as the orthogonality, which is used to characterise the complexity classes, is here shown to be related to the zeta function of the underlying dynamical systems.

\bibliographystyle{alphaurl}
\bibliography{Markov-zeta-lmcs-final}

\end{document}